
\newcount\mgnf\newcount\tipi\newcount\tipoformule\newcount\greco

\tipi=2          
\tipoformule=0   


\global\newcount\numsec
\global\newcount\numfor
\global\newcount\numtheo
\global\advance\numtheo by 1

\def\senondefinito#1{\expandafter\ifx\csname#1\endcsname\relax}

\def\SIA #1,#2,#3 {\senondefinito{#1#2}%
\expandafter\xdef\csname #1#2\endcsname{#3}\else
\write16{???? ma #1,#2 e' gia' stato definito !!!!} \fi}

\def\etichetta(#1){(\veroparagrafo.\veraformula)%
\SIA e,#1,(\veroparagrafo.\veraformula) %
\global\advance\numfor by 1%
\write15{\string\FU (#1){\equ(#1)}}%
\write16{ EQ #1 ==> \equ(#1) }}

\def\letichetta(#1){\veroparagrafo.\verotheo
\SIA e,#1,{\veroparagrafo.\verotheo}
\global\advance\numtheo by 1
\write15{\string\FU (#1){\equ(#1)}}
\write16{ Sta \equ(#1) == #1 }}

\def\tetichetta(#1){\veroparagrafo.\veraformula 
\SIA e,#1,{(\veroparagrafo.\veraformula)}
\global\advance\numfor by 1
\write15{\string\FU (#1){\equ(#1)}}
\write16{ tag #1 ==> \equ(#1)}}

\def\FU(#1)#2{\SIA fu,#1,#2 }

\def\etichettaa(#1){(A\veroparagrafo.\veraformula)%
\SIA e,#1,(A\veroparagrafo.\veraformula) %
\global\advance\numfor by 1%
\write15{\string\FU (#1){\equ(#1)}}%
\write16{ EQ #1 ==> \equ(#1) }}

\def\BOZZA{
\def\alato(##1){%
 {\rlap{\kern-\hsize\kern-1.4truecm{$\scriptstyle##1$}}}}%
\def\aolado(##1){%
 {
{
 \rlap{\kern-1.4truecm{$\scriptstyle##1$}}}}}
 }

\def\alato(#1){}
\def\aolado(#1){}

\def\veroparagrafo{\number\numsec}
\def\veraformula{\number\numfor}
\def\verotheo{\number\numtheo}

\def\Eq(#1){\eqno{\etichetta(#1)\alato(#1)}}
\def\eq(#1){\etichetta(#1)\alato(#1)}
\def\leq(#1){\leqno{\aolado(#1)\etichetta(#1)}}
\def\teq(#1){\tag{\aolado(#1)\tetichetta(#1)\alato(#1)}}
\def\Eqa(#1){\eqno{\etichettaa(#1)\alato(#1)}}
\def\eqa(#1){\etichettaa(#1)\alato(#1)}
\def\eqv(#1){\senondefinito{fu#1}$\clubsuit$#1
\write16{#1 non e' (ancora) definito}%
\else\csname fu#1\endcsname\fi}
\def\equ(#1){\senondefinito{e#1}\eqv(#1)\else\csname e#1\endcsname\fi}

\def\Lemma(#1){\aolado(#1)Lemma \letichetta(#1)}%
\def\Theorem(#1){{\aolado(#1)Theorem \letichetta(#1)}}%
\def\Proposition(#1){\aolado(#1){Proposition \letichetta(#1)}}%
\def\Corollary(#1){{\aolado(#1)Corollary \letichetta(#1)}}%
\def\Remark(#1){{\noindent\aolado(#1){\bf Remark \letichetta(#1).}}}%
\def\Definition(#1){{\noindent\aolado(#1){\bf Definition
\letichetta(#1)$\!\!$\hskip-1.6truemm}}}
\def\Example(#1){\aolado(#1) Example \letichetta(#1)$\!\!$\hskip-1.6truemm}

\def\include#1{
\openin13=#1.aux \ifeof13 \relax \else
\input #1.aux \closein13 \fi}

\openin14=\jobname.aux \ifeof14 \relax \else
\input \jobname.aux \closein14 \fi
\openout15=\jobname.aux

\let\EQ=\Eq


{\count255=\time\divide\count255 by 60 \xdef\hourmin{\number\count255}
        \multiply\count255 by-60\advance\count255 by\time
   \xdef\hourmin{\hourmin:\ifnum\count255<10 0\fi\the\count255}}

\def\oramin{\hourmin }

\def\data{\number\day/\ifcase\month\or january \or february \or march \or april
\or may \or june \or july \or august \or september
\or october \or november \or december \fi/\number\year;\ \oramin}

\newcount\pgn \pgn=1
\def\foglio{\number\numsec:\number\pgn
\global\advance\pgn by 1}
\def\foglioa{A\number\numsec:\number\pgn
\global\advance\pgn by 1}

\footline={\rlap{\hbox{\copy200}}\hss\tenrm\folio\hss}

\def\TIPIO{
\font\setterm=amr7 
\def \settepunti{\def\rm{\fam0\setterm}
\textfont0=\setterm   
\normalbaselineskip=9pt\normalbaselines\rm }\let\nota=\settepunti}

\def\TIPITOT{
\font\twelverm=cmr12
\font\twelvei=cmmi12
\font\twelvesy=cmsy10 scaled\magstep1
\font\twelveex=cmex10 scaled\magstep1
\font\twelveit=cmti12
\font\twelvett=cmtt12
\font\twelvebf=cmbx12
\font\twelvesl=cmsl12
\font\ninerm=cmr9
\font\ninesy=cmsy9
\font\eightrm=cmr8
\font\eighti=cmmi8
\font\eightsy=cmsy8
\font\eightbf=cmbx8
\font\eighttt=cmtt8
\font\eightsl=cmsl8
\font\eightit=cmti8
\font\sixrm=cmr6
\font\sixbf=cmbx6
\font\sixi=cmmi6
\font\sixsy=cmsy6
\font\twelvetruecmr=cmr10 scaled\magstep1
\font\twelvetruecmsy=cmsy10 scaled\magstep1
\font\tentruecmr=cmr10
\font\tentruecmsy=cmsy10
\font\eighttruecmr=cmr8
\font\eighttruecmsy=cmsy8
\font\seventruecmr=cmr7
\font\seventruecmsy=cmsy7
\font\sixtruecmr=cmr6
\font\sixtruecmsy=cmsy6
\font\fivetruecmr=cmr5
\font\fivetruecmsy=cmsy5
\textfont\truecmr=\tentruecmr
\scriptfont\truecmr=\seventruecmr
\scriptscriptfont\truecmr=\fivetruecmr
\textfont\truecmsy=\tentruecmsy
\scriptfont\truecmsy=\seventruecmsy
\scriptscriptfont\truecmr=\fivetruecmr
\scriptscriptfont\truecmsy=\fivetruecmsy
\def \eightpoint{\def\rm{\fam0\eightrm}
\textfont0=\eightrm \scriptfont0=\sixrm \scriptscriptfont0=\fiverm
\textfont1=\eighti \scriptfont1=\sixi   \scriptscriptfont1=\fivei
\textfont2=\eightsy \scriptfont2=\sixsy   \scriptscriptfont2=\fivesy
\textfont3=\tenex \scriptfont3=\tenex   \scriptscriptfont3=\tenex
\textfont\itfam=\eightit  \def\it{\fam\itfam\eightit}%
\textfont\slfam=\eightsl  \def\sl{\fam\slfam\eightsl}%
\textfont\ttfam=\eighttt  \def\tt{\fam\ttfam\eighttt}%
\textfont\bffam=\eightbf  \scriptfont\bffam=\sixbf
\scriptscriptfont\bffam=\fivebf  \def\bf{\fam\bffam\eightbf}%
\tt \ttglue=.5em plus.25em minus.15em
\setbox\strutbox=\hbox{\vrule height7pt depth2pt width0pt}%
\normalbaselineskip=9pt
\let\sc=\sixrm  \let\big=\eightbig  \normalbaselines\rm
\textfont\truecmr=\eighttruecmr
\scriptfont\truecmr=\sixtruecmr
\scriptscriptfont\truecmr=\fivetruecmr
\textfont\truecmsy=\eighttruecmsy
\scriptfont\truecmsy=\sixtruecmsy }\let\nota=\eightpoint}

\newfam\msbfam   
\newfam\truecmr  
\newfam\truecmsy 
\newskip\ttglue
\ifnum\tipi=0\TIPIO \else\ifnum\tipi=1 \TIPI\else \TIPITOT\fi\fi

\def\a{\alpha}
\def\b{\beta}
\def\d{\delta}
\def\e{\epsilon}

\def\g{\gamma}
\def\k{\kappa}
\def\l{\lambda}
\def\r{\rho}
\def\s{\sigma}
\def\t{\tau}

\def\z{\zeta}

\def\D{\Delta}
\def\L{\Lambda}
\def\G{\Gamma}

\def\E{{I\kern-.25em{E}}}
\def\N{{I\kern-.25em{N}}}
\def\M{{I\kern-.25em{M}}}
\def\R{{I\kern-.25em{R}}}
\def\Z{{Z\kern-.425em{Z}}}
\def\1{{1\kern-.25em\hbox{\rm I}}}
\def\eu{{1\kern-.25em\hbox{\sm I}}}

\def\C{{I\kern-.64em{C}}}
\def\P{{I\kern-.25em{P}}}
\def\eop{{ \vrule height7pt width7pt depth0pt}\par\bigskip}



\def\EE{{\cal E}}

\def\GG{{\cal G}}

\def\JJ{{\cal J}}

\def\LL{{\cal L}}

\def\SS{{\cal S}}
\def\TT{{\cal T}}

\def\MM{{\cal M}}

\def\LL{{\cal L}}

\def\RR{{\cal R}}
\def\QQ{{\cal Q}}

\def\chap #1#2{\line{\ch #1\hfill}\numsec=#2\numfor=1}

\def\sqr#1#2{{\vcenter{\vbox{\hrule height.#2pt
     \hbox{\vrule width.#2pt height#1pt \kern#1pt
   \vrule width.#2pt}\hrule height.#2pt}}}}


\newcount\foot
\foot=1
\def\note#1{\footnote{${}^{\number\foot}$}{\ftn #1}\advance\foot by 1}
\def\tag #1{\eqno{\hbox{\rm(#1)}}}
\def\frac#1#2{{#1\over #2}}

\def\text#1{\quad{\hbox{#1}}\quad}

\def\proof{{\noindent\pr Proof: }}

\def\remark{\noindent{\bf Remark: }}
\def\thanks{\noindent{\bf Aknowledgements: }}
\font\pr=cmbxsl10


\font\ch=cmbx12
\font\ftn=cmr8

\font\it=cmti10
\font\bf=cmbx10
\font\sm=cmr7

%
\catcode`\X=12\catcode`\@=11
\def\n@wcount{\alloc@0\count\countdef\insc@unt}
\def\n@wwrite{\alloc@7\write\chardef\sixt@@n}
\def\n@wread{\alloc@6\read\chardef\sixt@@n}
\def\crossrefs#1{\ifx\alltgs#1\let\tr@ce=\alltgs\else\def\tr@ce{#1,}\fi
   \n@wwrite\cit@tionsout\openout\cit@tionsout=\jobname.cit
   \write\cit@tionsout{\tr@ce}\expandafter\setfl@gs\tr@ce,}
\def\setfl@gs#1,{\def\@{#1}\ifx\@\empty\let\next=\relax
   \else\let\next=\setfl@gs\expandafter\xdef
   \csname#1tr@cetrue\endcsname{}\fi\next}
\newcount\sectno\sectno=0\newcount\subsectno\subsectno=0\def\r@s@t{\relax}
\def\resetall{\global\advance\sectno by 1\subsectno=0
  \gdef\firstpart{\number\sectno}\r@s@t}
\def\resetsub{\global\advance\subsectno by 1
   \gdef\firstpart{\number\sectno.\number\subsectno}\r@s@t}
\def\v@idline{\par}\def\firstpart{\number\sectno}
\def\l@c@l#1X{\firstpart.#1}\def\gl@b@l#1X{#1}\def\t@d@l#1X{{}}
\def\m@ketag#1#2{\expandafter\n@wcount\csname#2tagno\endcsname
     \csname#2tagno\endcsname=0\let\tail=\alltgs\xdef\alltgs{\tail#2,}%
  \ifx#1\l@c@l\let\tail=\r@s@t\xdef\r@s@t{\csname#2tagno\endcsname=0\tail}\fi
   \expandafter\gdef\csname#2cite\endcsname##1{\expandafter
     \ifx\csname#2tag##1\endcsname\relax?\else{\rm\csname#2tag##1\endcsname}\fi
    \expandafter\ifx\csname#2tr@cetrue\endcsname\relax\else
     \write\cit@tionsout{#2tag ##1 cited on page \folio.}\fi}%
   \expandafter\gdef\csname#2page\endcsname##1{\expandafter
     \ifx\csname#2page##1\endcsname\relax?\else\csname#2page##1\endcsname\fi
     \expandafter\ifx\csname#2tr@cetrue\endcsname\relax\else
     \write\cit@tionsout{#2tag ##1 cited on page \folio.}\fi}%
   \expandafter\gdef\csname#2tag\endcsname##1{\global\advance
     \csname#2tagno\endcsname by 1%
   \expandafter\ifx\csname#2check##1\endcsname\relax\else%
\fi
   \expandafter\xdef\csname#2check##1\endcsname{}%
   \expandafter\xdef\csname#2tag##1\endcsname
     {#1\number\csname#2tagno\endcsnameX}%
   \write\t@gsout{#2tag ##1 assigned number \csname#2tag##1\endcsname\space
      on page \number\count0.}%
   \csname#2tag##1\endcsname}}%
\def\m@kecs #1tag #2 assigned number #3 on page #4.%
   {\expandafter\gdef\csname#1tag#2\endcsname{#3}
   \expandafter\gdef\csname#1page#2\endcsname{#4}}
\def\re@der{\ifeof\t@gsin\let\next=\relax\else
    \read\t@gsin to\t@gline\ifx\t@gline\v@idline\else
    \expandafter\m@kecs \t@gline\fi\let \next=\re@der\fi\next}
\def\t@gs#1{\def\alltgs{}\m@ketag#1e\m@ketag#1s\m@ketag\t@d@l p
    \m@ketag\gl@b@l r \n@wread\t@gsin\openin\t@gsin=\jobname.tgs \re@der
    \closein\t@gsin\n@wwrite\t@gsout\openout\t@gsout=\jobname.tgs }
\outer\def\localtags{\t@gs\l@c@l}
\outer\def\globaltags{\t@gs\gl@b@l}
\outer\def\newlocaltag#1{\m@ketag\l@c@l{#1}}
\outer\def\newglobaltag#1{\m@ketag\gl@b@l{#1}}

\def\t@gsoff#1,{\def\@{#1}\ifx\@\empty\let\next=\relax\else\let\next=\t@gsoff
   \expandafter\gdef\csname#1cite\endcsname{\relax}
   \expandafter\gdef\csname#1page\endcsname##1{?}
   \expandafter\gdef\csname#1tag\endcsname{\relax}\fi\next}
\def\verbatimtags{\let\ift@gs=\iffalse\ifx\alltgs\relax\else
   \expandafter\t@gsoff\alltgs,\fi}
\catcode`\X=11 \catcode`\@=\active
\localtags
%
\setbox200\hbox{$\scriptscriptstyle \data $}
\global\newcount\numpunt
\hoffset=0.cm
\baselineskip=14pt
\parindent=12pt
\lineskip=4pt\lineskiplimit=0.1pt
\parskip=0.1pt plus1pt

\hyphenation{small}

\catcode`\@=11

\centerline{ \bf {PHASE SEPARATION FOR THE LONG RANGE ONE--DIMENSIONAL ISING MODEL
 \footnote*  {\eightrm
  This work
has been carried out thanks to the support of the A*MIDEX project (n$^{\rm o}$
ANR-11-IDEX-0001-02) funded by the "Investissements d'Avenir"  French
Government program, managed by the French National Research Agency (ANR).
}}}
\vskip.5cm
 \centerline{
Marzio Cassandro \footnote{$^1$}{\eightrm
Gran Sasso Science Institute, INFN Center for Advanced Studies
Viale F. Crispi 7, L'Aquila, 67100, Italy.
marzio.\-cassan\-dro@\-gmail.com},
Immacolata Merola  \footnote{$^2$}{\eightrm
Dipartimento di Matematica,
Universit\`a di L'Aquila -Via Vetoio, 1
67010 COPPITO (AQ), Italy.
immaco\-lata.me\-rola@\-dm.univaq.it},
Pierre Picco \footnote{$^3$}{\eightrm   Aix-Marseille Université,  CNRS,
Centrale Marseille,  I2M UMR 7373, 13453
Marseille France. pierre.picco@univ-amu.fr }
}

 \footnote{}{\eightrm {\eightit AMS 2000 Mathematics Subject Classification}:
Primary 60K35, secondary 82B20,82B43.}
\footnote{}{\eightrm {\eightit Key Words}:
Ferromagnetic Ising systems,
long range interaction,  phase transition, contours, Peierls estimates,
cluster expansion, phase segregation}

\vskip .5truecm

\centerline{ Dedicated to the memory of Enza Orlandi}
\vskip1cm
\noindent {\bf Abstract}
We consider  the phase separation  problem for the one--dimensional ferromagnetic Ising
model with long--range two--body interaction,  $J(n)=n^{-2+\a}$ where  $n\in \N$ denotes the distance of the two spins and  $ \alpha
\in ]0,\a_+[$  with $\a_+=(\log 3)/(\log 2) -1$.  We prove that
given $m\in ]-1,+1[$,
if the temperature is small enough,
then typical configuration for the $\mu^{+}$ Gibbs measure conditionally to have a empirical magnetization of the order $m$
are made of  a single interval that occupy  almost a proportion  $\frac{1}{2}(1-\frac{m}{m_\b})$ of the volume  with the minus
phase inside and the rest of the volume is the plus phase, here $m_\b>0 $ is the spontaneous magnetization.

\vskip1cm

 \chap { 1 Introduction and main results}1
\numsec= 1 \numfor= 1

We consider a  one--dimensional  ferromagnetic Ising model with a two body  interaction     $J(n)=n^{-2+\a}$
  where  $n$ denotes the distance of the two spins and  $ \alpha$
  tunes  the  decay of the
interaction.

A systematic and successful analysis of these one dimensional  models  started  more than forty years ago.
 [\rcite{GaMi},\rcite {Ru},\rcite{D0},\rcite{D1},\rcite{D2}] proved { existence},   uniqueness of the Gibbs states and
 analyticity in $\b$ of the free energy for $\a<0$ and [\rcite{Dy1},\rcite{Dy2},\-\rcite{Dy3}] proved the occurrence of a phase transition for $\a>0$.

The borderline case $\a=0$ was already distinguished by a number of unusual features
in the early seventies  [\rcite{Th}, \rcite{Dy3}].
It took  more than a decade to prove Dyson's conjecture [\rcite{Dy1}]  about the existence of a spontaneous magnetization
at low temperature. This result was proved by   Fr\"ohlich \& Spencer
[\rcite {FS}] by introducing a suitable notion of contours.
Precise estimates  on the low--temperature decay of the truncated correlations were  given  by Imbrie [\rcite{I}];
the existence of a Thouless effect [\rcite{Th}], that is a discontinuity of the magnetization at the critical temperature was
proved  by Aizenman, Chayes, Chayes \& Newman
[\rcite {ACCN}] and  all these works culminate
in  the proof of  the existence of an intermediate phase similar to a Kosterlitz-Thouless phase
with  a variable exponent  power law decay for correlation functions given by  Imbrie \& Newman  [\rcite{IN}].

{   In one--dimensional systems it was proved that  Fannes, Vanheuverzwijn \& Verbeure
 [\rcite{FVV}] and Burkov \& Sinai [\rcite{BS}] that used an energy argument that comes from Bricmont, Lebowitz \& Pfister
 [\rcite{BLP1}] there are no non-translation invariant extremal Gibbs states.  This argument was used in particular in
  [\rcite{BLP1}] to  prove  uniqueness of Gibbs states for one--dimensional Ising systems under the Ruelle's condition
 [\rcite{Ru}].}

In a more recent work,  [\rcite{CFMP}] revisited  [\rcite{FS}]  to extend  Peierls argument to the $0<\a<1$ case.
This  already allowed to study the behaviour of these systems when an external stochastic field is added [\rcite{COP1}, \rcite{COP2}]
and also to study the localization of the interface when taking, say $-$ boundary conditions on the left of an interval
and $+$ boundary on its right [\rcite{CMPR}].

Another important fact  currently observed in everyday life, therefore macroscopic,  is the  phase separation or phase  segregation phenomena.
To deduce it from Statistical Mechanics   was first considered in a remarkable paper on the short range Ising model by
Minlos \& Sinai in 1967, see [\rcite{MS}],  we share with  Pfister, see [\rcite{PF}], that "many important ideas, which were later on developed in Statistical Mechanics
were in germs in it". Then in 1988,
Dobrushin, Kotecky \& Shlosman, see  [\rcite{DKS}],  derived from statistical mechanics  the phenomenological
macroscopic theory of Wulff, that gives the  shape of the spatial region occupied by one phase immersed in the other one. Later  this  was called the
DKS theory.
See also Pfister [\rcite{PF}]  for recent version of Minlos \& Sinai work and of  DKS theory, also
 Pfister \& Velenik, see   [\rcite{PFVe1}] for large deviations and continuum limit  and  [\rcite{Io1}, \rcite{Io2}] for extensions of
 the DKS theory for all temperature below the critical one.

In this paper, we address the problem of phase separation or phase segregation   where the empirical magnetization is fixed in the interval
$]-m_\b,+m_\b[$, where $m_\b$ is the spontaneous magnetization. We assume that $\b$ is large
enough to have $m_\b>0$, a sufficient condition on $\b$ was given in [\rcite{CFMP}].

To be more precise, we consider in the finite interval $\L=[-L,+L]\cap \Z$, the system
of Ising spin configurations $\underline {\s}_{\L}=(\s_i, i\in \L)$  described by the Hamiltonian with $+$ boundary conditions
 $$
h^{++} (\s_{\L})=\frac{1}{2}\sum_{(i,j)\in \L\times \L} J(i-j)\frac{(1-\s_i\s_j)}{2}
+\sum_{i\in \L}\sum_{ j\in\L^c}J(i-j)\frac{(1-\s_i)}{2}.
\Eq(1)
$$
where the pair interaction $J(i-j)$ is defined by
 $$
 J(n)=\cases{ 0 & if $n=0$;\cr
 J+1 >>1& if $n=1$;\cr
 \frac 1 {|n|^{2- \alpha} } & if $n\neq 1$.\cr
 }\Eq (2)
$$
and $\alpha \in [0, \a_+ )$ with $\a_+=(\log 3)/(\log 2) - 1$.
In \eqv(2), the nearest neighbour interaction $J+1$ is chosen large enough as in [\rcite{CFMP}] appendix A, however new
conditions will be imposed here.
As proved in [\rcite{CFMP}], or in [\rcite{Dy1}] for such system, when $\b\ge \b_0(\a)$ for some $\b_0(\a)$ that comes from
an energy-entropy argument within a Peierls argument, there exist
at least two different extremal Gibbs states $\mu^+_\b$ and $\mu^-_\b$ that are limit when $|\L| \uparrow \infty$ of the finite
volume Gibbs measure with $+$, respectively $-$  boundary conditions. Then the spontaneous magnetization is
$$
m_\b=\mu^{+}_\b[\s_0]=-\mu^-_\b[\s_0]>0.
\Eq(emmebeta)
$$
 Let us take a $\b\ge \b_1(\a) (\ge\b_0(\a)) $ as in [\rcite{CMPR}] to have convergence of the cluster expansion then
it follows from theorem 2.5 in [\rcite{CMPR}] that
$$
\mu^+_\b [\s_0]= 1-\left[2 \xi^{++}(\b)\left(1+{ \cal B}(x,++)
\right)\right]
\Eq(preliminary)
$$
where
$$
\xi^{++}(\b)= e^{-2\b(\zeta(2-\a)+J)}
\Eq(ksipiu)
$$
with $\zeta(2-\a)$ the Riemann zeta function,
and  ${ \cal B}(x,++)$ is an absolutely convergent series that satisfies
 $$
 { |\cal B}(x,++)| \le e^{-\frac{\b}{32}(\frac{\z_\a}{\a(1-\a)}-3\d)}
 \Eq(est1triangle)
 $$
 where $\d$ is given in  \eqv(MP333) and $\z_\a=1-2(2^\a-1)$.

Let
$$
\e_0=|\L|^{-a}, \quad a>0
\Eq(epsilon00a)
$$
and given $m \in ]-1,+1[$, let
$$
\t=\frac{1-|m|}{2}
\Eq(tau1a)
$$
Let $\b^\star=\b^{\star}(|m|) $ such that $m_{\b^\star}=|m|+\t=(1+|m|)/2$.
Note that $|m|+\t< 1$ and  therefore $\b^\star <\infty$.
By GKS inequality we have :
for all $\b>\b^*$
$$
m_\b-|m|\ge m_{\b^*}-|m|=\t
\EQ(stac1a)
$$
we assume that $|\L|$ is large enough to have
$$
\t>\e_0
\EQ(tau1)
$$
therefore we get
 $$
\frac{m_\b-|m|}{m_\b} \ge \frac{\t}{m_\b}>\frac{\e_0}{m_\b}>\e_0.
\EQ(stac2b)
$$
\smallbreak
Let
$$
m_\L(\underline \s_\L)=\frac{1}{|\L|} \sum_{i\in \L} \s_i
\EQ(mag)
$$
be the empirical magnetisation.
We consider the system under the constraint that
$$
\left | m_\L(\underline\s_\L)
 -m \right| \le \e_0 m_\b.
\Eq(3)
$$
Note that $m_\L(\underline \s_\L) \in ]-m_\b,+m_\b[$, in fact
it follows from \eqv(3) and \eqv(tau1) that for all $\b > \b^\star$,  we have
$$
m_\b -\left | m_\L(\underline \s_\L) \right |>  \t-\e_0 m_\b>\t-\e_0 > 0
\EQ(stac3)
$$
{\it i.e.} $m_\L(\underline \s_\L)$ is well separated from $m_\b$ and $-m_\b$ as it should be.

Since the interaction is ferromagnetic, under the constraint \eqv(3) with  $\e_0=0$  and $m$ is a rational number
of the form $k/|\L|$ for some positive odd integer $k\le |\L|$,  say $k=2q+1$,
a minimum of \eqv(1) is reached by a configuration made of a single run of ${-1}$ of length
$L-q$.
In other word a ground state
contains a single interval of $-1$ with the correct length to satisfies \eqv(3) with $\e_0=0$. However
it can be located anywhere
in $\L$, therefore the ground state is $L+1+q$ times degenerated.
The main problem is therefore to understand what remains of this $\b=\infty$ picture for the configurations that are
typical for the Gibbs measure $\mu^{+}_\b$ conditioned by \eqv(3) when we take $\b$ large enough but finite.

Roughly speaking, we show that for $|\L|$ very large,  the  configurations that are typical for the finite volume conditional $\mu^{++}_
\L$ measure, given \eqv(3), are as follows:

\noindent {\bf 1} There exists in $\L$ an interval $\L^\prime$  with $|\L^\prime|\approx (|\L| \frac{1}{2}(1-\frac{m}{m_\b}) $
where
$$
\frac{1}{|\L^\prime|} \sum_{i\in \L^\prime} \s_i\approx -m_\b;
\Eq(4)
$$
while
$$
\frac{1}{|\L\setminus \L^\prime|} \sum_{i\in \L\setminus \L^\prime} \s_i\approx m_\b.
\Eq(41)
$$

\noindent{\bf 2} The statistics of the spin configurations, in the limit $|\L|\uparrow \infty$, in the interval  $\L^\prime$ are
similar to the one of the
$-$ phase, while in the region $\L\setminus \L^\prime$ they are similar  the one of the $+$ phase.

\medskip

In this paper we deal only with the case $\alpha>0$.
For  $\alpha=0$ the argument goes along the same lines but  now , since   many of the bounds we use  are no more exponential,  the proofs
require  substantial modifications.
These proofs will be presented in our next paper where we study the fluctuations of the interval where , with plus boundary conditions, the phase is negative.
 In this case the value of alfa should play a relevant role. In fact  from our previous work [\rcite{CMPR}] , where we study the separation point when the right and left  boundary conditions are different $(+,-)$  we have:
 1) for  $\alpha>0$  the density distribution of  the fluctuations of the transition point ,suitably normalized, converges to a Gaussian ,
 2) for  $\alpha=0$,rescaling the volume $ [-L,L]$ to $[-1,1]$,  the distribution converges to a non degenerate distribution with an explicit density in the interval $[-1,1]$.

\medskip

In section 2, we define the model, state the main theorem  and two propositions that will imply the theorem.

In section 3, we state and prove two  lemmata  that will be extendedly used.

In section 4 and 5 we give the proofs of the two propositions of section 2.

All the proofs in these sections  are based on the geometrical description of
spin configurations introduced in [\rcite{CFMP}]
and use the cluster expansion developed in [\rcite{CMPR}].
The appendix contains a resum\'e of the results obtained in [\rcite{CFMP}], [\rcite{CMPR}].

\medbreak
 \chap { 2 Definitions and main result }2
\numsec= 2 \numfor= 1
\numtheo=1

Let $\L=[-L,+L]\cap \Z$ and $\SS_\L=\{-1,+1\}^\L$ be  the set of spin configurations in $\L$. We denote by
$\underline \s_\L\equiv(\s_i, i\in \L)\in \SS_\L$ a
configuration. For any subset $A\subset \L$, we denote by $\underline \s_A\equiv(\s_i, i\in A)$ the restriction
of the configuration $\underline \s_\L$
to the subset $A$.
For $f$ a cylindrical bounded function with cylinder basis  that is a subset of $\L$,  the finite volume Gibbs
measure with $+$ boundary conditions is given by
$$
\mu^{+}_\L[f]=\frac{\sum_{\underline \s_\L\in \SS_\L} f(\underline \s_\L) e^{-\b h^{++} (\underline \s_{\L})}}{Z_\L^{++}(\b)}
\Eq(21)
$$
The infinite volume limit, $\lim_{\L\uparrow \Z} \mu^{+}_\L(f)$ exists by FKG inequalities, see  [\rcite{E}],   [\rcite{FKG}] or [\rcite{OV}] and
is translation invariant as all extremal Gibbs state are by [\rcite{BS}] and [\rcite{FVV}].

\smallbreak
\Definition(S0)
$\,\,${\it For $\e_0=|\L|^{-a}$, with $0<a<1$ to be chosen later,  assuming that $|\L|$ is large enough to have  \eqv(tau1), and  for $m\in [-1,+1]$, let}
$$
\SS_\L(m,\e_0)=\left\{ \underline \s_\L \in \SS_\L\,:\, |m_\L(\underline \s_\L ) -m | < \e_0 m_\b \right\}.
\Eq(esse01)
$$

Since the boundary condition is fixed, there is a bijection between  spin configurations and
spin-flip configurations, a spin-flip being  a  pair of consecutive sites $(i,i+1)$ with
$\s_i\s_{i+1}=-1$. Triangles are a graphical representation of  pairing together
spin-flips, say $(i,{i+1})$ and $(j, {j+1}) $  where $i<j$,  with the property that
$\s_{i+1}\s_j=+1$. It is  obtained by an algorithm described in [\rcite{CFMP}], see also section 6.
In particular two triangles are either disjoint or one inside the other.

The mass of a triangle T will be denoted by $|T|$ and is just the number of sites of $\Z$ in the base of the triangle,
{\it i.e} if $T$ is associated to the two spin-flips  $(i,i+1)$ and $(j,j+1)$ with $i<j$  then $|T|=j-i$.
\smallbreak
 We say that a family of triangles $\underline T$ is compatible, if there exists a
 spin configuration
$\underline \s_\L$ such that $\underline T=\underline T(\underline \s_\L)$, this spin
configuration will be denoted by $\underline \s_\L(\underline T)$.
The set of compatible configurations of triangles will be denoted by $\TT_\L$.

\smallbreak

\medskip

\Definition(Ssmall)
$\,\,$ {\it For $\e_s=|\L|^{-\g}$  with $0<\g<1$ to be chosen later, let
$$
\TT_\L^{small}(\e_s)= \left\{ \underline T \in \TT_\L  \,:\, \forall T\in \underline T, \, |T|\le  \e_s|\L|\right\}
\Eq(smalltriangles)
$$
be the set of compatible  configurations of small triangles and
$$
\SS_\L^{small}(\e_s)= \left\{ \underline \s_\L\in \SS_\L \,:\,  \underline T(\underline \s_\L) \in \TT_\L^{small}(\e_s) \right\}
\Eq(small)
$$
be the set of spin configurations such that the associated family of triangles is made of small triangles.}

\medskip

\Remark(sc) {\it It follows from} [\rcite{CMPR}] {\it that for $\e_s=|\L|^{-\g}$, there exists a $\b_0(\a)$ such that for all $\b\ge \b_0(\a)$,
the typical configurations of triangles for the
measure $\mu^{+}_\L$ are within $\SS_\L^{small}(\e_s)$, in the sense that $\lim_{\L\uparrow \Z} \mu^{+}_\L[\SS_\L^{small}(\e_s)]=1$.
 The set $\SS_\L^{small}(\e_s)$ plays the  r\^ole of the phase of small contours
as in} [\rcite{PF}].
\medskip
\Definition(exter) $\,\,${\it
For a given family of compatible  triangles $\underline T\in \TT_\L$,  we say that a triangle $T\in \underline T$ is external
with respect to $\underline T$ or more simply,  external,  if it is not contained in any other triangle of $\underline T$.
We say that a family of triangles $\underline T$ is made of  mutually external triangles if each triangle $T\in \underline T$  in external
with respect to any other triangle of $\underline T$ }

\smallbreak

\Definition(Largeext) $\,\,${\it
Given a configuration $\underline \s_\L\in \SS_\L$, we denote by $\underline T^{E} (\underline \s_\L)$ the subfamily
of all triangles in $\underline T(\underline \s_\L)$ that are external with respect to $\underline T(\underline \s_\L)$ and not small.} {\it i.e.}
$$
\underline T^{E} (\underline \s_\L)=\left\{ T\in \underline T(\underline \s_\L)\, :\, T\, {\rm is \, external}, \,|T|> \e_s |\L|\right\}
\Eq(cext)
$$
{\it  On the other hand, given a family $\underline T^E$ of mutually external triangles that satisfies $\forall T\in \underline T^E$, $|T|>\e_s|\L|$
we say that it is a compatible family if there exists a configuration $\underline \s_\L$  such that
$\underline T^E=\underline T^E(\underline \s_\L)$.}

\smallbreak

\Definition(Text)$\,\,\,${\it
Given $\e_s>0$ let
$$
\TT_\L^E=\left\{ \underline T \in \TT_\L \,:\,
\underline T {\rm \,are\, mutually\, external}, \forall \widetilde T \in \underline T, |\widetilde T|>\e_s|\L|\right\}.
\Eq(calText)
$$
}

If $\underline T^E \in \TT_\L^E$ we denote
$$
|\underline T^E|=\sum_{T \in \underline T^E} |T|.
\EQ(massText)
$$

Given $\underline T^E\in \TT^E_\L$, there is a specific spin configuration, say $\overline \s(\underline T^E)$,  defined by
$$
\overline \s_i(\underline T^E)=-\1_{\{i\in \D(\underline T^E)\}}+ \1_{\{i\in \L\setminus \D(\underline T^E)\}}
\Eq(simplest)
$$
where $\D(\underline T^E)$ is just the union of the bases over all the triangles, large and external, that define  $\underline T^E$.

\smallbreak
Let us define an equivalence relation on $\SS_\L$ by $\underline \s_\L\sim \underline \s^\prime_\L$
if and only if $\underline T^E(\underline \s_\L)=\underline T^E(\underline \s^{\prime}_\L)$.

\smallbreak
\Definition(eqcla)
{ \it Given a $\underline T^E \in \TT^E_\L$,
let $S_{\underline T^E}$ be the equivalent class of spin configurations corresponding to the representative
$\underline T^E$:
$$
S_{\underline T^E}\equiv\{\underline \s_\L\in \SS_\L: \underline T^E(\underline \s_\L)=\underline T^E\}.
\EQ(essegae)
$$
}
\smallbreak
Then we have $S_{\underline T_1^E} \neq S_{\underline T_2^E}$ if $\underline T^E_1\neq \underline T_2^E$ and
therefore, recalling \eqv(small)  the partition
$$
S_\L=\SS_\L^{small}(\e_s) \bigcup_{\underline T^E\in \TT^E_\L} S_{\underline T^E}.
\Eq(parti)
$$

Using  \eqv(simplest), the point is that $\forall\, \underline \s_\L\in S_{\underline T^E}$ with
$\underline \s_\L \neq  \overline \s(\underline T^E)$ we have
$$
h^{++}(\underline \s_\L) > h^{++} (\overline \s(\underline T^E)).
\Eq(GS)
$$

Notice that  given  $ \underline \s_\L \in \SS_{\underline T^E}$,
all triangles $ \widetilde T \in \underline T(\underline \s_\L)\setminus \underline T^E(\underline \s_\L)$ describe
fluctuations with respect to the fondamental state $\underline \s(\underline T^E)$.

\smallbreak
\Definition(SE=)
$\,\,$ {\it Given $0< \rho\le 1$, for
$ \e_s =|\L|^{-\g}$ with $0<\g < 1$, and assume that $|\L|$ is large enough to have $\r\ge \e_s$,   let
$$
\TT^E_\L(\r) =\left\{ \underline T^E \in \TT^E_\L : |\underline T^E|=\r|\L|\right\}
\Eq(calTrho)
$$
and
$$
\SS_\L^1(\r)
= \Big\{ \underline \s_\L \in \SS_\L\,:\,  \underline T ^E(\underline \s_\L) \in  \TT^E_\L(\r) \Big\}.
\Eq(se=)
$$
}
\smallbreak
\Remark (notation) {\it
In this article the letter
$\TT$ always denotes  a set of compatible configuration of triangles  as  for $\TT^E_\L(\r)$ in \eqv(calTrho)
while the letter ${\SS}$ always denotes a set of spin configurations as for $\SS_\L^1(\r)$ in \eqv(se=). }
 \smallbreak

Note that $\SS_\L^1(\r)$ depends on $\e_s$ since $\e_s$ appears in the definition of $\TT^E$, however for simplicity
we do not write this dependence.
Moreover  if $0<\r< 1$ is not a rational number that can be written as $k/|\L|$ for some integer $k$, the set $\SS_\L^1(\r)$ is empty.
For future use let us denote
$$
Q_\L=\left\{ \r\in [0,1] \,:\, \exists k \in \{0,\dots,|\L|\}\,, \, \r=k/|\L|\right\}.
\Eq(queue)
$$
\smallbreak
\noindent {\bf \Lemma(entropy)} {$\,\,$} Given $0<\r\le 1$, and $\e_s=|\L|^{-\g}$,  the number
of configurations of external triangles in $\TT^E_\L(\r)$, say  $\sharp  [\TT^E_\L(\r)]$,  satisfies
$$
 \sharp  [\TT^E_\L(\r)]\le e^{(2-\g) |\L|^\g \log |\L|}.
\EQ(entro)
$$
The proof is done  in section 6.
\smallbreak

 For  $\e_c=|\L|^{-\nu}$, $0<\nu<1$,  we define also
 $$
\SS^1_\L(\r,\e_c)=\bigcup_{\r-\e_c\le \r^\prime\le \r+\e_c} \SS_\L^1(\r^\prime).
\EQ(esseuno)
$$
Note that the previous union is merely over the set $\{(\r-\e_c)\vee \e_s  \le \r^\prime\le (\r+\e_c)\wedge 1\} \cap Q_\L$ whose
 cardinality is less  than $|\L|$.

\smallbreak
\Definition(esseB)
$\,\,${\it For all $\r \in (0,1)\cap Q_\L$ and for $\e_c=|\L|^{-\nu}$ and $|\L|$ large enough to have  $8\e_c < \r$,
for all $\underline T^E\in  \TT_\L^E(\r) $
let
$$
n_0[\underline T^E]=\sum_{T^E\in \underline T^E} \1_{\{|T^E|\ge |\underline T^E|-6\e_c|\L|\}}
\Eq(enn0g)
$$
be the number of triangles in $\underline T^E$ with mass larger than $|\underline T^E|-6\e_c|\L| $, a number larger than $2\e_c|\L|$.
Moreover let
$$
\TT^B_\L(\r)=\left\{ \underline T^E \in \TT^E_\L(\r)\, :\, n_0[\underline T^E]=1\right\}
\EQ(tibiro)
$$
and
 $$
\SS_\L^B(\r) =\left\{ \underline \s_\L\in \SS_\L^1(\r)
\, : \, \underline T^E(\underline \s_L) \in \TT^B_\L(\r) \right\}
\Eq(esseb)
$$
be the set of spin configurations that give rise to a family of external triangles of total mass $|\underline T^E|=\r |\L|$ that
contains  a single external triangle, say $T_1$, that satisfies
$|T_1|\ge (\r-6\e_c)|\L|$. }
\smallbreak

\Remark (others)
{\it Notice  that the total mass of the other  triangles of $\underline T^E$, that are mutually
external and large by definition,  fits the rest of the volume
{\it i.e.}
$$
\sum_{T\in \underline T^E, T\ne T_1} |T| =\r|\L|-|T_1| \le 6\e_c |\L|.
\Eq(otherlarge)
$$
}
\noindent In a way similar to \eqv(esseuno) and with the same convention,  we define
$$
\SS_\L^B(\r,\e_c)=
\bigcup_{\r-\e_c\le \r^\prime\le \r+\e_c} \SS_\L^B(\r^\prime).
\Eq(essebuono)
$$

\smallbreak
\noindent{\bf Choice of the parameters and some rounding conditions}

 Given $m\in]-1,1[$, and assume as in the introduction that $\b>\b^\star$ to have  \eqv(stac2b), let us define
 $$
 \hat \r(m)=\frac{1}{2}\left(1-\frac{m}{m_\b}\right).
 \Eq(hatrho)
 $$
 The point is that we have the following
 \medskip
  \noindent{\bf  \Lemma(rho)}
 {\it If $\b>\b_1(\a)$,
 for all $\widetilde m\in ]-1,1[$,  for all $\underline T^E\in \TT^E_\L$ such that $|\underline T^E|=\hat\rho(m)|\L|$ then
$$\mu_\L[m_\L(\underline \s_\L)|\SS_{\underline T^E}]=m\,\pm \,\frac{10\xi^{++}(\b)}{\a(1-\a)} \frac{1}{|\L|^{1-\a}}
 \Eq(fondarho)
 $$
 where $\xi^{++}(\b)$ is defined in \eqv(ksipiu).
On the other hand, for all $\r\in ]0,1[$, for all
all $\underline T^E\in \TT^E_\L$ such that $|\underline T^E|=\rho|\L|$ then
$$\mu_\L[m_\L(\underline \s_\L)|\SS_{\underline T^E}]=(1-2\rho)m_\b\,\pm \,\frac{10\xi^{++}(\b)}{\a(1-\a)} \frac{1}{|\L|^{1-\a}}
 \Eq(fondarhobis)
 $$
}
  The proof which is a consequence of  the cluster expansion will be done in section 6.

 Note that it follows from \eqv(stac2b) that
 $$
 \frac{\t}{2m_\b} \le \hat \r(m) \le 1-\frac{\t}{2m_\b}.
 \Eq(stac4)
 $$
 To avoid rounding problems, let us define
$$
\r(m)\equiv \r_\L(m)=\sup\{\r\in Q_\L :\, \r\le \hat \r(m)\}
\EQ(rhoinv)
$$
 where $Q_\L$ is defined in \eqv(queue). Notice that $\r(m)\le \hat\r(m)$ and $\hat \r(m)-\r(m) \le |\L|^{-1}$.

Now we collect conditions on the parameters introduced above.
We always assume that
$$
\e_s=|\L|^{-\g} \, , \,\e_0=|\L|^{-a} \,\, , \,{\rm and}\,\, \e_c=|\L|^{-\nu}.
\EQ(epsmallzero)
$$
with
$$
0<\g<\min\{ \a-\nu, 2/3\}  \, ;\,
\frac{ \g +\nu \a}{1-\a} \le
(1-\nu)\a \,\,; \,\, \nu <a\,\, {\rm and}\,\,
\nu < \g(1-\a).
\Eq(delga)
$$

\Remark (PT)  {\it The first condition
$0<\g<\a-\nu$ comes from \eqv(condextra1234),
the condition $0<\g<2/3$ is stated before \eqv(onlyone).
The condition $\frac{ \g +\nu \a}{1-\a} \le
(1-\nu)\a $  comes from \eqv(extracond1),
to be able to find an $\eta$ in between $\frac{ \g +\nu \a}{1-\a} $ and $
(1-\nu)\a$,  $\,$
$\nu <a$  is for \eqv(main2) where $2\e_c -\e_0$ is present and the last
$\nu < \g(1-\a)$
 condition is \eqv(condextra123)  {\it i.e.}
$\a+\g(1-\a)-\nu >\a, $
}

\Remark( PT1)
{\it It is easy to check that  a possible choice is
$$
\nu= \frac{\a(1-\a)}{4} \, ; \,\,\g=\frac{\a}{4} \,;\,\, a=\frac{\a(1-\a)}{2}.
\Eq(provchoice)
$$
}

\smallbreak
The following theorem shows how the phenomenon of phase separation holds for
long range Ising model in one dimension at low temperature.
It is the analogue of the Minlos \& Sinai  theorem and its extension by Dobrushin, Kotecky \& Shlosman
that hold for the two dimensional short range Ising model.
\smallbreak

\noindent {\bf \Theorem(mainalt)}
$\,\,$ {\it
For all $0<\a<\a_+$,
 for $m\in ]-1,1[$,
for $(\e_0,\e_s, \e_c )$ that satisfy \eqv(epsmallzero) and \eqv(delga)
there exists a $\b_{3}(\a,m )$ such that
for all $\b\ge \b_{3}(\a, m)$
we have
$$
\lim_{\L\uparrow \Z} \mu^{+}_\L\big[\SS_\L^B(\r(m),\e_c)|S_\L(m,\e_0) \big]=1.
\Eq(main1alt)
$$
 }

\medbreak

The proof of the theorem \eqv(mainalt) is a direct consequence of the two following propositions
that give more precise estimates
for the  involved probabilities.
\medskip
\noindent{\bf \Proposition(miming)} $\,$
{\it For all $0<\a<\a_+$,
for $m\in ]-1,1[$,
for $(\e_0,\e_s, \e_c )$ that satisfy \eqv(epsmallzero) and \eqv(delga)
there exists a $\b_4(\a,m )
$ such that
for all $\b\ge \b_4(\a,m)$, if $|\L|$ is large enough, we have
$$
\mu^{+}_\L\big[\SS_\L^1(\r(m),\e_c)|S_\L(m,\e_0) \big]\ge
1- \left[ e^{-\frac{\z_\a }{16} \b(2\e_c-\e_0)m_\b|\L|^{\a+\g(1-\a)}  }+
e^{-\frac{\b (\r(m) |\L|)^\a}{2\a(1-\a)} \left\{(1+\frac{\e_c}{\r(m)})^\a-1
\right\} }.
\right]
\Eq(main2)
$$
where $\z_\a=1-2(2^\a-1)$.}

\medskip

\noindent{\bf \Proposition(miming2)} $\,$
{\it For all $0<\a<\tilde \a_+$,
for $m\in ]-1,1[$,
for $(\e_0,\e_s, \e_c )$ that satisfy \eqv(epsmallzero) and \eqv(delga)
there exists a $\b_{5}(\a,m)$
such that
for all $\b\ge \b_{5}(\a,m)$, if $|\L|$ is large enough,  we have}
$$
  \mu^{+}_\L\big[\SS_\L^1(\r(m),\e_c)|S_\L(m,\e_0) \big]=
  \mu^{+}_\L\big[\SS_\L^B(\r(m),\e_c)|S_\L(m,\e_0) \big]\times
   \left[ 1\pm  \frac{ e^{-\frac{\b}{\a(1-\a)} \frac{\z_\a}{4} (\e_c|\L|)^\a}
}{1-2e^{-\frac{\b}{\a(1-\a)} \frac{\e_0}{\r(m)}  c(\a) \{\r(m) |\L|\}^\a}}\right]
   \Eq(main3)
$$
{\it where $c(\a)$ is given in \eqv(calpha)}.

The proof of proposition \eqv(miming2) will be done in section 4, the one of proposition \eqv(miming) in section 5.

\medbreak
 \chap { 3 Preparatory lemmata}3
\numsec= 3 \numfor= 1
\numtheo=1

In this section we first give an estimate from below of the energy cost to fragmentize a large triangle.
Then we give an estimate for the Laplace transform of the probability distribution of the empirical magnetization for the Gibbs
measure conditioned to some
specific subsets of configurations.

 Since the system is ferromagnetic and the strength of the interaction between two spins decays
 with their distance,  conditionally on $S_\L(m,\e_0)$,
 one can expect that at low temperature the system prefers to  form
 a single interval of consecutive minuses
(therefore a triangle) of size of order $\r(m) |\L|$
instead of various intervals
whose sum of lengths will be of order $\r(m) |\L|$.
 We first show that it is true at the level of the energy.

 So let us start with $\{I_1, I_2, I_3,\dots,I_\k\}$ a family of disjoint intervals in $\L$, labelled in such a way that $I_i$ is on the left of $I_{i+1}$  for all
$i\in\{1,\dots,k-1\}$ and
 let $\underline \s_\L(I_1,\dots,I_k)$ be a spin configuration where $\underline \s_\L(I_1,\dots, I_k)$ is minus
  in each of the intervals $I_j$ and plus on $\L\setminus \cup_{j=1}^{k}I_j$, then we have
 $$
 h^{++}(\underline \s_\L(I_1,\dots,I_k)) =\sum_{i=1}^{k} h^{++} (\s_{ I_i}\circ 1_{\L\setminus I_i}) -2\sum_{1\le i<j \le k} W(I_i,I_j)
 \EQ(decomp1)
 $$
 where  $\s_{ I_i}\circ 1_{\L\setminus I_i}$ is the configuration which is minus on $I_j$ and plus on $\L\setminus I_i$ and
 $$
 W(I_i,I_j)=\sum_{\ell_1 \in I_i}\sum_{\ell_2\in I_j} \frac{1}{|\ell_1-\ell_2|^{2-\a}}\, .
 \EQ(interac)
 $$
 Let $d(i,j)$ be the inter-distance between $I_i$ and $I_{j}$, since $W(I_i,I_j) $
 is a decreasing function of $d(i,j)$ any transformation that decreases some $d(i,j)$ and keeps constant the others
 decreases the  energy \eqv(decomp1).
 In particular if we set all the $d(i,j)$ equal to zero, that is we merge all the intervals $\{I_1, I_2, I_3,\dots,I_\k\}$ in a single one, say
 $I^*_{1,\dots,k}$  then we have
$$
h^{++}(\underline \s_\L(I_1,\dots,I_k))- h^{++}(\underline \s_\L(I^*_{1,\dots,k}))>0.
\EQ(reara3)
$$
In the following Lemma we exploit the structure of triangles  to get
 an explicit lower bound for this difference in some specific cases.
 \smallskip
 Given
$\r \in Q_\L$, see \eqv(queue)
and $\e_c$ that satisfies  \eqv(epsmallzero) and \eqv(delga)
we can assume that $|\L|$ is large enough to have $\r \ge 8\e_c$,
recalling \eqv(calTrho),
let us define
$$
\TT_\L^B(\r)\equiv\{ \underline T^E\in \TT_\L^E(\rho)  \,:\, \sup_{T\in \underline T^E} |T| \ge (\r-6\e_c) |\L|\} .
\EQ(extrbrho)
$$
Note that with this definition, recalling \eqv(esseb) and remark \eqv(others), we have
$\SS^B_\L(\r)=\{\underline \s_\L \in \SS_\L : \underline T^E(\underline \s_\L)\in \TT_\L^B(\r)\}$.

We introduce  a discretisation of an interval of size $2\e_c$ around a generic point $\r$:
$$
B(\r,\e_c)=[\r-\e_c,\r+\e_c]\cap Q_\L,
\EQ(bro)
$$
in the special case where $\r=\r(m)$ defined in \eqv(rhoinv), for simplicity we denote
$$
B(m, \e_c)=B(\r(m), \e_c).
\EQ(bm)
$$

\noindent{\bf \Lemma(rie)} $\,$ {\it
For all $\r \in B(m,\e_c)$, with $\e_c$ that satisfies  \eqv(epsmallzero) and \eqv(delga), for all
$\underline T^E\in  \TT_\L^E(\r)\setminus \TT_\L^B(\r)$, let $T_0$ be  an arbitrary  triangle with $|T_0|=\r(m) |\L|$,
recalling the definition of $\overline{\s}(\underline T^E)$ in  \eqv(simplest), we have
$$
h^{++}(\overline{\s}(\underline T^E))-h^{++}(\overline{\s}(T_0))\ge \frac{\z_\a}{2\a(1-\a)} (\e_c|\L|)^\a
\EQ(quantriesz)
$$
where $\z_\a=1-2(2^\a-1)$ which is strictly positive if $\a<\a_+=(\log3)/(\log2) -1$.
 }

\medskip
\proof

  Given an interval $\JJ\subset \Z$, and $\underline T^E \in \TT_\L^E$ let
  $$
  {\cal Q}(\JJ,\underline T^E)=|\JJ\cap \D(\underline T^E)|
  \EQ(charge)
  $$
 be the number of points in  $\D(\underline T^E)=\cup_{T\in \underline T^E} \D(T)$  that belong to $\JJ$.
 Recall that $\underline T^E$ is made of mutually external triangles, therefore the previous union is over
 disjoint triangles.
 Note that $ {\cal Q}(\JJ,\underline T^E)\le |\JJ|$
 and  ${\cal Q}(\JJ,\underline T^E)$ is additive with respect to $\JJ$, that is
  if $\JJ=\JJ_1\cup\JJ_2$ with $\JJ_1\cap \JJ_2=\emptyset$  we have ${\cal Q}(\JJ,\underline T^E)={\cal Q}(\JJ_1,\underline T^E)+
  {\cal Q}(\JJ_2,\underline T^E)$.

 Let $T_1$ be a    triangle in $\underline T^E$  with the largest mass.
 Since $\underline T^E \in \TT_\L^E(\r)\setminus \TT_\L^B(\r)$, we have $|T_1| <(\r-6\e_c)|\L|$.

 Let us first assume that $|T_1|\ge 2\e_c |\L|$, the other case will be treated later.
Let $I^>
 =\{x \in \Z\,:\, x>\D(T_1)\}$ and $I^<
  =\{x\in \Z\,:\, x<\D(T_1)\}$ be the semi-infinite intervals respectively on the right and on the left of $T_1$.
  Since $\sum_{T\in \underline T^E} |T|=\r|\L|$ we have
 $$
  {\cal Q}(I^>,\underline T^E)+ {\cal Q}(I^<,\underline T^E)=\sum_{T\in \underline T^E} \1_{\{T\ne T_1\}} |T|=\r|\L| -|T_1|\ge 6\e_c|\L|
  \EQ(R1)
  $$
 then either   ${\cal Q}(I^>,\underline T^E)$ or  ${\cal Q}(I^<,\underline T^E)$ should be larger than $3\e_c|\L|$, let us assume that it is
 ${\cal Q}(I^>,\underline T^E)$.
 Given a positive number   $\k<3\e_c$, to be fixed later,   let $I^>_\k
  =\{x\in \Z\,:\, x-\k |\L| >\D(T_1)\}$ be the semi-infinite interval
 at distance $\k|\L|$ of $T_1$ and on its right.
 Let $I^+_\k=\{x\in \Z\,:\, x>\D(T_1), x-\k|\L|<\D(T_1)\}$, the  interval of length $\k|\L|$ on the right of $T_1$ and adjacent to it.
 Since ${\cal Q} (I^+_\k, \underline T^E) \le \k |\L|$, by additivity with respect to $\JJ$ of $ {\cal Q} (\JJ,   \underline T^E)$,  we have
 recalling that by hypothesis ${\cal Q}(I^>,\G^E)\ge 3\e_c|\L|$,
 $$
  {\cal Q}(I^>_\k
  ,\underline T^E)\ge (3\e_c-\k)|\L|.
  \EQ(pete)
  $$

 We define a new configuration ${\Theta} (\overline \s(\underline T^E))$ as follows:
 \item{1)} All the spins $\s_i$ with $i\in I^+_\k$ that are minus are changed in plus. This will erase all the triangles in $\underline T^E$ that
 have basis  in $I^+_\k$.
 If a triangle, say $\tilde T_1$ in $\underline T^E$  intersect both $I^+_\k$ and $I^>_\k$ then it becomes a smaller triangle with base
 $\D(\tilde T)\cap I^>_\k$.

 \item{2)} We merge all the bases of triangles $\tilde T$ such that $\D(\tilde T) \subset I^>_\k$ into a single interval,
 say $I^\star_2$,  that we put at a distance exactly $\k |\L|$ from $T_1$, the length of this interval is
 $$
 |I^\star_2|={\cal Q}(I^>,\underline T^E)\ge (3\e_c-\k)|\L|
  \EQ(pete2)
$$
 and all the spin are minus there.

 \item{3)} We merge all the bases of the triangles $\hat T$ such that $\D(\hat T) \subset I^<(T_1)$ with $\D(T_1)$ to
 get an interval, say $I^\star_1$
 where all the spins are minus.
 Using that $|T_1|\ge 2\e_c |\L|$, we have
 $$ |I^\star_1|\ge 2\e_c|\L|.
 \Eq(pete3)
 $$

 We get a spin configuration, $\Theta(\overline {\s} (\underline T^E))$ which is minus on  $I^\star_1 \cup I^\star_2$ and
 plus everywhere else. We have
 $ {\rm dist} (I^\star_1, I^\star_2) =\k |\L|$ with \eqv(pete2) and \eqv(pete3). If we choose $\k=\e_c$ we get
 $$
 {\rm dist}(I^\star_1, I^\star_2)=\e_c|\L| \le \min\{|I^\star_1|, |I^\star_2| \}.
 \Eq(anti)
 $$
 Recalling Definition \eqv(triangles), the family of triangles $\underline T({\Theta} (\overline \s(\underline T^E)))$
 cannot be made of two triangles with basis
 $I^\star_1$ and $ I^\star_2$
 as it  follows comparing
 \eqv(anti) and   \eqv(prop:dist-T).

 Therefore it should be  made of a large triangle with basis the interval $[x_-(I^\star_1), x_+(I^\star_2)]$, say $T^{\star\star}_1$
 with $|T^{\star\star}_1|=|I^\star_1|+\e_c|\L|+|I^\star_2|$.  We have  $|I^\star_1|+|I^\star_2|\ge (\r-\e_c)|\L|$ because
 we erase less than $\e_c|\L|$ minuses
 in the volume $I^+_\k$ with $\k=\e_c$.  Therefore we get
 $$
 |T^{\star\star}_1|\ge \r|\L|.
 \EQ(anti2)
 $$
  Inside this triangle $T^{\star\star}_1$, there is a triangle, say $T^{\star\star}_2$, of size exactly $\e_c|\L|$.
 We have
 $$
 h^{++}(\overline{\s}(\underline T^E))-h^{++}(\Theta(\overline {\s} (\underline T^E)))\ge 0.
 \EQ(attrac)
 $$
 To get \eqv(quantriesz) it remains to estimate from below $h^{++}(\Theta(\overline {\s} (\underline T^E)))-h^{++}(\overline \s(T_0))$.

 It is precisely here that we use the structure of triangles. Using  (2.4), (2.6) and (2.8) in [\rcite{CFMP}],
 noticing that $T^{\star\star}_2$ is the smallest triangle and
 taking  into account of the missing $\a(1-\a)$, see after Lemma \eqv(tr1), we get
 $$
 h^{++}(\Theta(\overline {\s} (\underline T^E)))-h^{++}(\overline \s(T_0))\ge
 \frac{\z_\a}{\a(1-\a)}(\e_c |\L|)^\a  +h^{++}(T^{\star\star}_1)-h^{++}(\overline \s(T_0)).
 \EQ(riesz1)
 $$
Since for a configuration made of a  single triangle $T$ in $\Z$ we have
 $$
 \frac{2|T|^\a}{\a(1-\a)} -\frac{2}{\a} \le h^{++}(\overline \s(T))\le  \frac{2|T|^\a}{\a(1-\a)}  -2\left(1-\frac{1}{\a}\right),
 \EQ(basic)
 $$
 we get, for $|\L|$ sufficiently large and how large depends on $m,\a,$ and $\nu$, see \eqv(epsmallzero),
 $$
 \eqalign{  h^{++}(\Theta(\overline {\s} (\underline T^E)))-h^{++}(\overline \s(T_0))
  &\ge
   \frac{\z_\a}{\a(1-\a)}(\e_c |\L|)^\a  +\frac{2}{\a(1-\a)}\left[( \r(m)-\e_c)^\a-(\r(m))^\a\right] |\L|^\a
  -2\cr
&\ge   \frac{\z_\a}{2\a(1-\a)}(\e_c |\L|)^\a
}   \EQ(basic2)
  $$
 where we used that $\r>\r(m) -\e_c$.

 It remains to consider the case $|T_1|< 2\e_c |\L|$, then all the triangles of $\underline T^E$ have a mass
 that is smaller than $2\e_c|\L|$, since by definition
 $|T_1|$ is a triangle with maximal mass in $\underline T^E$.

  Given an integer $t\in [x_-(\underline T^E), x_+(\underline T^E)]$, let $\QQ(t,\underline T^E)=\QQ([[x_-(\underline T^E)], t],\underline T^E)$
  where $[x_-(\underline T^E)]$ is
  the integer part of $x_-(\underline T^E)$.  $\QQ(t,\underline T^E)$ is  the number site $i\le t$ where  $\s_i=-1$ in $\overline{\s}(\underline T^E)$.
   Then, $\QQ(t,\underline T^E)$
  is a non-strictly  increasing function that increases linearly with a slope 1 when $t\in \D(T)$ for some $T\in \underline T^E$ and is constant
  for $t$ in between two such  triangles. We have also $\QQ([x_-(\underline T^E)],\underline T^E)=0$ and
  $\QQ([x_+(\underline T^E)],\underline T^E)=\r|\L|$. In particular
   the graph of this function intersects the level $\r|\L|/2$ in two possible ways: either in a constant part of the graph or in the
   linear part of the graph.
So let $p=\inf\{p\in \Z\,:\,  \QQ(p,\underline T^E)=\r|\L|/2\}$. Let $\JJ^+ $ be an interval centered at $ p$ and of size $\e_c|\L|$, then
$\QQ(\JJ^+,\underline T^E)\le \e_c|\L|$.
On the other hand
$\QQ([x_-(\underline T^E)], p-\frac{\e_c}{2}|\L|], \underline T^E)\ge (\frac{\r}{2} -\e_c) |\L|$ and
$\QQ([p+\frac{\e_c|\L|}{2}, x_+(\underline T^E)], \underline T^E) \ge   (\frac{\r}{2} -\e_c)|\L|$.
At this point we proceed as before: all the spin $\s_i$ with $i\in \JJ^+$ that are minus are changed in plus ;
we merge all the bases
of the triangles that are on the left of $\JJ^+$ and all the triangles that are on the right of $\JJ^+$.
The rest of the proof is the same as above. \eop

\smallbreak

Now we collect some estimates for the Laplace transform of the probability distribution of the empirical
 magnetization conditioned on two kind of subsets
of spin configurations.
The first kind of subsets are simply $\SS_{T_0}$ with $T_0$ a triangle with $|T_0|=\r|\L|$, see \eqv(essegae).
For the other ones,
recalling \eqv(se=),
given $\underline T^E, \in \TT_\L^E(\rho) $, see \eqv(calTrho), let
$$
\SS^{VS}_\L(\underline T^E,\r,\e_s ) =\left\{\underline \s_\L \in \SS_{\underline T^E} \cap S^1_\L(\r) \,:\, \forall
\widetilde T \in \left\{ \underline T(\s_\L)\setminus  \underline T^E \right\},
\,\, |\widetilde T|\le \e_s |\L|
\right\}
\EQ(titti120pr)
$$
be the set of spin configurations that give rise to the  family $\underline T^E$, but  all the  associated triangles
that are not in $\underline T^E$ are small. This is relevant for the triangles that are internal to $\underline T^E$.

\noindent{\bf \Lemma(lpc)} $\,$ {\it
There exists a $\b_{6}(\a,m)$ such that for all $\b\ge \b_{6}(\a,m)$,
for all $t$ such that
$$
|t| \le
\frac{\z_\a }{4\a(1-\a)(\e_s |\L|)^{1-\a}}
\EQ(largestt)
$$
where $\z_\a=1-2(2^\a-1)$ and  $\e_s$  is defined in \eqv(epsmallzero),
we have
$$
\left| \log \mu^+_\L \left[e^{\b t\sum_{i\in \L} \s_i} \big | \SS^{VS}_\L(\underline T^E,\r,\e_s )\right] -
 \b t  |\L|\, \mu_\L^{+}[m_\L(\underline \s_\L)|\SS^{VS}_\L(\underline T^E,\r,\e_s )]
  \right|\le
\frac{\b^2 }{2} t^2|\L| e^{-2\b J}.
\EQ(lpc1a)
$$
On the other hand for all $\r\in ]0,1] $, for all $t$ such that
$$
|t|\le
\frac{\zeta_\a3^{1-\a}}{4\a(1-\a) (\r |\L|)^{1-\a}}
\EQ(smallestt)
$$
for all $T_0$ with $|T_0|=\r |L|$, we have
$$
\left| \log \mu^+_\L \left[e^{\b t\sum_{i\in \L} \s_i} \big | \SS_{T_0} \right] -
 \b t  |\L|\, \mu_\L^{+}[m_\L(\underline \s_\L)|\SS_{T_0}]
  \right|\le
\frac{\b^2 }{2} t^2|\L| e^{-2\b J}.
\EQ(lpc1)
$$
}

\proof

It follows from the Taylor formula that for any set of configurations $\widetilde \SS$ we have
$$
\log \mu^+_\L \left[e^{\b t\sum_{i\in \L} \s_i} \big | \widetilde \SS \right] -
 \b t  |\L|\, \mu_\L^{+}[m_\L(\underline \s_\L)|\widetilde \SS]=
\b^2 |\L|
\int_{0}^{t}\,ds\int_0^{s} \,dr \frac{1}{|\L|}\sum_{(i,j)\in \L\times \L} \mu^{+}_\L(r)[\s_i,\s_j\big | \widetilde \SS]
\Eq(5bcebis)
$$
where for all $i,j \in \L$
$$
\mu^{+}_\L(r)[\s_i,\s_j| \widetilde \SS]= \mu^{+}_\L(r)[\s_i\s_j| \widetilde \SS]- \mu^{+}_\L(r)[\s_i| \widetilde \SS]\mu^{+}_\L(r)[\s_j| \widetilde \SS]
\EQ(covarb)
$$
and for any cylindrical  function $f$ and $r\in \R$
$$
\mu_\L^{+}(r)[f| \widetilde \SS]=
\frac{ \sum_{\underline \s_\L\in \widetilde \SS} f(\underline \s_\L) e^{-\b h^{++}(\underline\s_\L)} e^{\b r \sum_{\in \L}\s_i}
}
{\sum_{\underline \s_\L\in \widetilde \SS}e^{-\b h^{++}(\underline\s_\L)}
}.
\EQ(condbtilde)
$$

We need an estimate uniform in $|r|\le t$ of the two point truncated correlation function \eqv(covarb) for the conditioned measure
\eqv(condbtilde) with a magnetic field $r$. Note that in all the considered cases the magnetic field is going to zero when $|\L|\uparrow \infty$.

This will be done using the cluster expansion.
As shown in chapter 6, a sufficient condition to be able to use the cluster expansion of [\rcite{CMPR}] in presence
of a magnetic field $t$ is simply
$$
\frac{\z_\a}{2\a(1-\a)} |\widetilde T|^\a -|t||\widetilde T| \ge \frac{\z_\a}{4\a(1-\a)} |\widetilde T|^\a
\EQ(condsuf)
$$
{\it i.e.}
$$
|t| \le  \frac{\z_\a}{4\a(1-\a)} \inf_{\widetilde T \in \widetilde \SS} \frac{1}{|\widetilde T|^{1-\a}}
\EQ(altcondsuf)
$$
In the case $\widetilde \SS=  \SS^{VS}_\L(\underline T^E,\r,\e_s )$ this gives
\eqv(largestt) while
in the case $\widetilde \SS=   \SS_{T_0}$ with  $|T_0|=\r |L|$, this give
\eqv(smallestt) where we have used
\eqv(unterzo6) that implies
that all such triangles $\widetilde T$  that are internal to  some triangle $ T^\star(\widetilde T)  \in \underline T^E$ satisfies
$$
|\widetilde T| \le \frac{1}{3} |T^\star(\widetilde T) |\le \frac{1}{3} |\underline T^E|.
\Eq(unterzo)
$$
\eop


\medbreak
 \chap { 4  Proof of the Proposition  \eqv(miming2)}4
\numsec= 4 \numfor= 1
\numtheo=1

 \smallskip

 Let us first note that since  $ \SS_\L^1(\r(m),\e_c) \supset
  \SS_\L^B(\r(m),\e_c)$ we have
 $$ \mu^{+}_\L\big[\SS_\L^1(\r(m),\e_c)|S_\L(m,\e_0) \big]\ge
  \mu^{+}_\L\big[\SS_\L^B(\r(m),\e_c)|S_\L(m,\e_0) \big].
  \EQ(trivial1)
  $$
  Therefore we need just to prove an  upper  for $ \mu^{+}_\L\big[\SS_\L^1(\r(m),\e_c)|S_\L(m,\e_0) \big]$.
 We start with
  $$
\eqalign{
 \mu^{+}_\L\big[\SS_\L^1(\r(m),\e_c)|S_\L(m,\e_0) \big] =
 &  \mu^{+}_\L\big[\SS_\L^B(\r(m),\e_c)|S_\L(m,\e_0) \big]
 \cr
 & \times  \left[ 1+ \frac{\mu^{+}_\L\big[\SS_\L^1(\r(m),\e_c)\setminus \SS_\L^B(\r(m),\e_c)
 ,S_\L(m,\e_0) \big] }
{ \mu^{+}_\L\big[\SS_\L^B(\r(m),\e_c),S_\L(m,\e_0) \big]\ }\right].
  }\Eq(mars1)
$$
where $\mu^{+}_\L[A,B]=\mu^{+}_\L[A\cap B]$.
 Recalling \eqv(tibiro),  let us define   for $\r\in ]0.1[$ and $x\in \L$,
 $$
 \TT^{E\setminus B}_{\L}(\r,x)=\left\{ \underline T^E\in \TT^E_\L(\r)\setminus \TT^B_\L(\r)\,;\, x_-(\underline T ^E)=x\right\}
 \Eq(geaicomp) $$
 where, see \eqv(def:delta-T),
 $ x_-(\underline T ^E)=\min_{T\in \underline T^E} x_-(T)$.

 We have
$$
\eqalign{
&\mu^{+}_\L\big[\SS_\L^1(\r(m),\e_c)\setminus \SS_\L^B(\r(m),\e_c) ,S_\L(m,\e_0) \big]
\le \cr
&\quad\quad\quad\quad\quad\frac{1}{Z^{++}_\L} \sum_{x\in \L}\,
\sum_{\r \in B(m,\e_c)}\, \sum_{\underline T^E\in  \TT^{E\setminus B}_{\L}(\r,x)}\,\,
\sum_{\underline \s_\L \in \SS_{\underline T^E}\cap S_\L(m,\e_0)} e^{-\b h^{++}(\underline \s_\L)}
 }\EQ(step1)
 $$
 where $ B(m,\e_c)$ is defined in \eqv(bm) and $ \SS_{\G^E}$ is defined in \eqv(essegae).

 On the other hand, for  $T_0$  a fixed triangle with basis  in $\L$ with mass $|T_0|=\r(m)|\L|$,
 recalling \eqv(se=),   we denote
 $$
 \SS_{T_0}(\r(m) )=\left\{\underline \s_\L\in S^1_\L(\r(m)) \, ;\, \underline T^E(\underline \s_\L)=T_0\, \right\}.
 \EQ(esse123)
 $$
 Then we have
 $$
 \mu^{+}_\L\big[\SS_\L^B(\r(m),\e_c),S_\L(m,\e_0) \big]
 \ge \frac{1}{Z^{++}_\L} \sum_{\underline \s_\L \in \SS_{T_0}(\r(m))
 \cap S_\L(m,\e_0)}  e^{-\b h^{++}(\underline \s_\L)}.
 \Eq(denom)
 $$
For $\underline T^E\in
  \TT^{E\setminus B}_{\L}(\r,x)
  $ with $\r \in B(m,\e_c)$,
  recalling the definition of $\overline\s(\underline T^E)$ in
 \eqv(simplest),  for simplicity let us denote
 $$
 \widetilde Z^{++}_\L(\underline T^E,m,\e_0)=\sum_{\s_\L\in \SS_{\underline T^E}\cap
 \SS(m,\e_0)}e^{-\b[h^{++}(\underline \s_\L)-h^{++}(\overline\s(\underline T^E))]}
 \Eq(partren)
 $$
even if $Z^{++}_\L(\SS_{\underline T^E}, \SS(m,\e_0))$ should be less ambiguous.
$ \widetilde Z^{++}_\L(T_0,m,\e_0)$ is defined  analogously when $\underline T^E=T_0$.
To avoid confusion, note that  in this constrained partition function
not only the triangle $T_0$ is present but it is the largest external triangle and all the other external triangles are small.

 Calling ${\cal R}_1(T_0) $ the ratio of the right hand side of \eqv(step1) over the right hand side of \eqv(denom), we have

 $$
 {\cal R}_1(T_0)\le
 \sum_{x\in \L}\, \sum_{\r \in B(m,\e_c)}\,\,\sum_{\underline T^E\in  \TT^{E\setminus B}_{\L}(\r,x)}\,
 e^{-\b[h^{++}(\overline \s(\underline T^E)-h^{++}(\overline \s(T_0))]}
 \frac{ \widetilde Z^{++}_\L(\underline T^E,m,\e_0)}{ \widetilde Z^{++}_\L(T_0,m,\e_0)}.
 \Eq(step2)
  $$

It remains to consider the last ratio in \eqv(step2).
Let us denote for $\underline T^E\in \TT^E_\L$ an arbitrary family of external triangles,
$$
\widetilde Z^{++}_\L(\underline T^E)=\sum_{\s_\L\in S_{\underline T^E}}e^{-\b[h^{++}(\underline \s_\L)-h^{++}(\overline\s(\underline T^E))]}
\EQ(zeegamma)
$$
 then we have $\widetilde Z^{++} _\L(\underline T^E,m,\e_0)\le \widetilde Z^{++} _\L(\underline T^E)$.
 We estimate  from below $\widetilde Z^{++}_\L(T_0,m,\e_0)$, when $T_0$ satisfies $|T_0|=\r(m)|\L|$.
 We claim that there exists a $\b_7(\a)$ such that for all $\b\ge \b_7(\a)$, we have
 $$
 \widetilde Z^{++} _\L(T_0,m,\e_0)\ge \widetilde Z^{++} _\L(T_0) \left(1-2e^{-\frac{\b}{\a(1-\a)} \frac{\e_0}{\r(m)} c(\a) \{\r(m) |\L|\}^\a}\right)
 \Eq(claim1)
 $$
 where
 $$
 c(\a)=
 \frac{\z_\a3^{1-\a} m_\b}{16}
  \EQ(calpha)
 $$
 from which we get
 $$
 {\cal R}_2\equiv \frac{ \widetilde Z^{++}_\L(\underline T^E,m,\e_0)}{ \widetilde Z^{++}_\L(T_0,m,\e_0)}
 \le \frac{\widetilde Z^{++}_\L(\underline T^E)}{\widetilde Z^{++}_\L(T_0)}
 \frac{1}{1-2e^{-\frac{\b}{\a(1-\a)} \frac{\e_0}{\r(m)}  c(\a)\{\r(m)|\L|\}^\a}}\equiv{\cal R}_3
 \frac{1}{1-2e^{-\frac{\b}{\a(1-\a)} \frac{\e_0}{\r(m)} c(\a) \{\r(m)|\L|\}^\a}}
 \EQ(postclaim1)
 $$

 Let us first prove the claim \eqv(claim1) and then  estimate ${\cal R}_3$.
 The claim will be a consequence of an estimate for
 $$
\frac{ \widetilde Z^{++}_\L(T_0,m,\e_0)}{\widetilde Z^{++}_\L(T_0)}
 =\mu^{+}_\L\left[S_\L(m,\e_0)|S_{T_0}\right]
 =1- \mu^{+}\left[\MM^>_\L(m,\e_0)|S_{T_0}\right]
 -\mu^{+}\left[\MM^<_\L(m,\e_0)|S_{T_0}\right]
  \EQ(postclaim2)
  $$
   where
  $$
  \eqalign{
  \MM^>_\L(m,\e_0)&:=\{m_\L(\underline \s_\L)>m+\e_0m_\b\};\cr
  \MM^<_\L(m,\e_0)&:=\{m_\L(\underline \s_\L)<m-\e_0m_\b\}.
  }
  \Eq(Esse><)
  $$
Therefore it is  enough to get an upper bound for the last two terms in \eqv(postclaim2).
 Let us consider first
    $\mu^{+}\left[\MM^>_\L(m,\e_0)|S_{T_0}\right]$. By Markov inequality,
  we have for $t_1>0$
  $$
    \mu^{+}\left[\MM^>_\L(m,\e_0)|S_{T_0}\right]\le e^{-\b t_1 |\L|(m+\e_0m_\b)} \mu^{+}_\L\left[e^{\b t_1 \sum_{x\in \L} \s_i}|S_{T_0}\right].
  \EQ(postclaim3)
  $$
Using Lemma \eqv(lpc) and Lemma \eqv(rho) to estimate $\mu_\L^{++}[m_\L(\underline \s_\L)|\SS_{T_0}]$ and
assuming
$$
\frac{\e_0}{2} m_\b \ge
\frac{10\xi^{++}(\b)}{\a(1-\a)} \frac{1}{|\L|^{1-\a}}
+ \frac{\b^2 }{2} t_1 e^{-2\b J}.
\EQ(condespilonzero2)
$$
we get, for $t$ satisfying \eqv(smallestt) with $\rho=\r(m)$,
$$
 \mu^{+}\left[\MM^>_\L(m,\e_0)|S_{T_0}\right]\le e^{-\b t_1 |\L|\frac{\e_0}{2} m_\b}.
\EQ(finalinbeauty)
$$

Let us now consider the second term in the right hand side  of \eqv(postclaim2), by Markov inequality for any $t_2\geq 0$, we have
$$
\mu^{+}\left[
\MM^<_\L(m,\e_0)
|S_{T_0}\right]\le
e^{+\b t_2 |\L|(m-\e_0m_\b)} \mu^{+}_\L\left[e^{-\b t_2 \sum_{x\in \L} \s_i}|S_{T_0}\right]
  \EQ(postclaim3bis)
 $$
the estimates are similar  to the previous ones   and under the same condition  \eqv(condespilonzero2) for $t_2$  instead of $t_1$ we get
$$
 \mu^{+}\left[S^>_\L(m,\e_0)|S_{T_0}\right]\le e^{-\b t_2 |\L|\frac{\e_0}{2} m_\b}
\EQ(finalinbeautymeno)
$$
This end the proof of the claim \eqv(claim1).
\medskip

Recalling \eqv(postclaim1), it remains to estimate ${\cal R}_3={\widetilde Z^{++}_\L(\underline T^E)}/{\widetilde Z^{++}_\L(T_0)}$
for $\underline T^E\in {\cal T}^{E\setminus B}_\L(\r,x)$.
This will be done using a  cluster expansion. By  Lemma \eqv(Lfond), used for $t=0$, we consider
first the leading terms of $\log {\cal R}_3$ which is
$$
\EE(\underline T^E, T_0)
=\sum_{x\in \L}\left( \xi^{\overline \s(\underline T^E)}(x) - \xi^{\overline \s(T_0)}(x)\right)
\Eq(leading)
$$
where
$$
\xi^{\overline \s(\underline T^E)}(x) =
\cases{e^{-\b \left ( 2J + \sum_{y\in \underline T^E \setminus \{x\}}J(x-y)\right)},
& if  $x\in \D(\underline T^E)\setminus {\rm sf}(\underline T^E)$;\cr
e^{-\b\left( 2J+  \sum_{y\in \L\setminus \D(\underline T^E) } J(x-y)\right) },
& if  $x\notin \D(\underline T^E) \cup  {\rm sf}(\underline T^E)$;\cr
}
\Eq(xi123)
$$
and a similar formula holds for $\xi^{\overline \s(T_0)}(x)$. Note first that $0\le \xi^{\overline \s(\underline T^E)}(x)\le 1$ and
$0\le \xi^{\overline \s(T_0)}(x)\le 1$.

Given $0<\eta<1$ to be chosen later,  let us denote
$$
\partial_\eta T_0=\{x\in \L \,:\, {\rm dist}(x,T_0)\le |\L|^\eta\}
\EQ(parttzero)
$$
and
$$
\partial_\eta \underline T^E =\{x\in \L \,:\, {\rm dist} (x,\underline T^E)\le |\L|^\eta\}.
\EQ(parttext)
$$
It is easy to check that
$$
|\partial_\eta T_0| =4|\L|^\eta \,\, {\rm and} \,\, |\partial_\eta \underline T^E| \le 4|\L|^\eta \sharp[\underline T^E]
\Eq(cardinal)
$$
where $\sharp[\underline T^E]$ is the number of triangles in the family $\underline T^E$.
Note that since $\underline T^E\in \cup_{x\in \L} {\cal T}^{E\setminus B}_\L(\r,x)$ and all triangles
in $\underline T^E$ are larger than $\e_s|\L|$, where
$\e_s$ satisfies \eqv(epsmallzero), we have
$$
\sharp[\underline T^E]\le \frac{\r |\L|}{\e_s|\L|} \le |\L|^\g.
\EQ(cardinal2)
$$
Therefore
$$
\sum_{x\in \partial_\eta T_0\cup \partial_\eta \underline T^E } \left| \xi^{\overline \s(\underline T^E)}(x) - \xi^{\overline \s(T_0)}(x)\right|
\le
8 |\L|^{\g+\eta} e^{-2J\b}.
\Eq(val1)
$$
To estimate the remaining terms in \eqv(leading), recalling \eqv(ksipiu),  we write them as
$$
\EE^{1} (\underline T^E, T_0)=
\xi^{++}(\b) \sum_{ x\in \L\setminus(\partial_\eta T_0\cup \partial_\eta \underline T^E )}
\left[\frac{\xi^{\overline \s(\underline T^E)}(x)}{\xi^{++}(\b)}-
\frac{\xi^{\overline \s(T_0)}(x)}{\xi^{++}(\b)}\right].
\Eq(remain)
$$
Using \eqv(xi123) and \eqv(ksipiu), the two ratios  in the bracket in \eqv(remain) are larger than 1.
However, given $x \in \L\setminus(\partial_\eta T_0\cup \partial_\eta \underline T^E )$, by comparison with an integral we have
$$
\sum_{ y\in \L} \frac{ \1_{\{{\rm dist}(x,y)> |\L|^\eta\}}}{|x-y|^{2-\a}} \le \frac{2}{|\L|^{\eta(1-\a)} } \frac{1}{1-\a}\equiv 2 g(|\L|)
\Eq(inte)
$$
Using that for $0\le z \le 2 g(|\L|) $, if $|\L|$ is large enough to have $2g(|\L|) \le 1$ then
$$
1+z \le e^z\le 1+z\left(1+\frac{z}{2} e^z\right) \le 1+z\big(1+3g(|\L|)\big)
\EQ(tay)
$$
we get, if $|\L|$ is large enough and how large depends on $\b$ and $(\a,\eta)$,
$$
\sum_{ x\in \L\setminus(\partial_\eta T_0\cup \partial_\eta \underline T^E )}
\frac{\xi^{\overline \s(\underline T^E)}(x)}{\xi^{++}(\b)}
\le |\L| + 2\b \sum_{x\in \D(\underline T^E)}\sum_{y\in \L\setminus \D(\underline T^E)} J(x-y) \big (1+3g(|\L|)\big)
\EQ(up)
$$
and
$$
\sum_{ x\in \L\setminus(\partial_\eta T_0\cup \partial_\eta \underline T^E )}
\frac{\xi^{\overline \s(T_0)}(x)}{\xi^{++}(\b)}
\ge |\L| -\left(\frac{6|\L|^\eta}{1-\a}\right) +2 \b \sum_{x\in \D(T_0)} \sum_{y \in \L\setminus \D(T_0)} J(x-y)
\EQ(down)
$$
Therefore, collecting \eqv(val1), \eqv(up) and \eqv(down),   we get
$$
\EE(\underline T^E, T_0)
\le
2\b \xi^{++}(\b) \left[ h^{++} (\overline \s(\underline T^E)) -h^{++}(\overline \s(T_0))\right]  + \xi^{++}(\b)\left(\frac{6|\L|^\eta}{1-\a}\right)
+ \xi^{++}(\b)\frac{6}{\a(1-\a)^2} |\L|^{\a+\g -(1-\a)\eta}
$$
where the last term correspond to $g(|\L|)$ in \eqv(up) and comes from the following rough estimate:
$$
h^{++}(\overline \s(\underline T^E)) 
\le \sum_{\widetilde T\in \underline T^E} h^{++}(\widetilde T)
\le \frac{ 2(\r|\L|)^\a}{\a(1-\a)} \sum_{\widetilde T\in \underline T^E} 1
\le \frac{ 2(\r|\L|)^\a}{\a(1-\a)} \frac{\r}{\e_s} \le \frac{2 \r^{1+\a} |\L|^{\a+\g}}{\a(1-\a)}
\Eq(newone)
$$
and \eqv(inte).

Since we want to use the lower bound \eqv(quantriesz)  and $(\e_c|\L|)^\a=|\L|^{(1-\nu)\a}$,
we assume
$$
\eta \le (1-\nu)\a \,\, {\rm and}\,\, \a+\g-\eta(1-\a) \le (1-\nu)\a
\EQ(extracond)
$$
that is
$$
\frac{ \g +\nu \a}{1-\a} \le \eta \le (1-\nu)\a .
\EQ(extracond1)
$$

Adding the error terms of the cluster expansion that are of the form $
 (1\pm e^{-\frac{\b}{32} (\frac{\zeta_\a}{\a(1-\a)}-3\d)})
$ to $\log {\cal R}_3$,
 inserting the result in \eqv(step2), using Lemma \eqv(rie) and Lemma \eqv(entropy)
we get that there exists a
$\b_5=\b_5(\a)$ such that for all $\b\ge \b_5(\a)$, if $|\L|$ is large enough,  we have
$$
{\cal R}_1 (T_0) \le \frac{ e^{-\b\frac{\z_\a}{4\a(1-\a)}(\e_c|\L|)^\a}
}{1-2e^{-\frac{\b}{\a(1-\a)}\frac{\e_0}{\r(m)} c(\a) \{\r(m)|\L|\}^\a}}.
\EQ(step317)
$$

Therefore, collecting \eqv(trivial1),  \eqv(mars1), \eqv(step2) and \eqv(step317) we get
$$
 \mu^{+}_\L\big[\SS_\L^1(\r(m),\e_c)|S_\L(m,\e_0) \big] =
 \mu^{+}_\L\big[\SS_\L^B(\r(m),\e_c)|S_\L(m,\e_0) \big] \,\left[ 1\pm  \frac{ e^{-\b\frac{\z_\a}{4\a(1-\a)}(\e_c|\L|)^\a}
}{1-2e^{-\frac{\b}{\a(1-\a)}\frac{\e_0}{\r(m)}  c(\a) \{\r(m) |\L|\}^\a}}\right]
\Eq(step318)
$$
this prove \eqv(main3). \eop

\medskip


 \medbreak
 \chap { 5  Proof of the Proposition  \eqv(miming)}5
\numsec= 5 \numfor= 1
\numtheo=1

\medskip

Let us first  give  a lower bound for probability of the event we are conditioning on:

{\bf \noindent \Lemma(lb132)}
{\it There exists a $\b_{8}=\b_{8} (\a)$ such that for all $\b\ge \b_8$, if $|\L|$
is large enough, we have
 $$
 \mu^{+}_\L[\SS_\L(m,\e_0)]\ge
 e^{-\frac{2\b}{\a(1-\a)} (\r(m)|\L|)^\a[1-\l_-(\b)]}
 (1-2\eta_1(\b,\L))
  \Eq(stimabasso)
 $$
where
$$
\l_-(\b)=2\xi^{++} (\b)(1-e^{-\frac{\b}{32} (\zeta_\a-3\d)})
\Eq(lambdameno)
$$
with
$\xi^{++}(\b)$ defined in \eqv(ksipiu), $\z_\a$ and $ \d$ defined in Lemma \eqv(Lfond), and
 $$
 \eta_1(\b,\L)= e^{-\frac{\b}{8} \e_0 m_\b \z_\a |\L|^{\a+\g(1-\a)}}.
 \Eq(etauno)
 $$ }

 \noindent{ \bf Proof}

 Given a triangle $T$ with $|T|=\r(m) |\L|$ and $\D(T)\subset \L$, recalling \eqv(essegae), \eqv(se=), and
using   \eqv(titti120pr) in the particular case $\underline T^E=T$, let
  $$
 \SS^{VS}_\L(T, \r(m),\e_s)=\{ \underline \s_\L\in \SS_T\cap \SS^1_\L(\r(m)) \,:\,
 \, \forall \widetilde T  \neq T, |\widetilde T |<\e_s|\L|\}.
 \EQ(verysmallT)
 $$
 We have
 $$
 \mu^{+}_\L[\SS_\L(m,\e_0)]\ge\left[1-\mu_\L^{++}[(\SS_\L(m,\e_0))^c|\SS^{VS}_\L(T, \r(m),\e_s)]\right] \mu^{+}_\L[ \SS^{VS}_\L(T, \r(m),\e_s)].
\Eq(reduc)
 $$
 Let us start with a  lower bound for $\mu^{+}_\L[ \SS^{VS}_\L(\r(m),\e_s)]$.
  Using  \eqv(simplest) with $\underline T^E=T$ and
   $|T|=\r(m)|\L|$, we have

 $$
  \mu^{+}_\L[ \SS^{VS}_\L(\r(m),\e_s)]=\frac{1}{Z_\L^{++}(\b)}
   e^{-\b h^{++}(\overline \s_T)} \sum_{\underline \s_\L\in  \SS^{VS}_\L(T, \r(m),\e_s)}
  e^{-\b[h^{++}(\underline \s_\L)-h^{++}(\overline \s_T)]}
  \Eq(oneterm)
 $$
using  the cluster expansion to estimate
$$
-\log Z_\L^{++}(\b)+
\log \big[\sum_{\underline \s_\L\in  \SS^{VS}_\L(T, \r(m),\e_s)}
  e^{-\b[h^{++}(\underline \s_\L)-h^{++}(\overline \s_T)]}\big]
\Eq(inter)
$$
we get
$$
 \eqalign{
  \mu^{+}_\L[ \SS^{VS}_\L(\r(m),\e_s)]&\ge
  e^{-\b h^{++}(\overline \s_T)}
    e^{\xi^{++} (\b)  2\b
 h^{++}(\overline\s_T) (1-e^{-\frac{\b}{32} (\zeta_\a-3\d)})}\cr
&\ge
e^{-\b h^{++}(\overline \s_T)[1-\l_-(\b)]}\cr
}\EQ(twoterm)
$$
where $\l_-(\b)$ is defined in \eqv(lambdameno).  Using \eqv(basic) we get
$$
 \mu^{+}_\L[ \SS^{VS}_\L(\r(m),\e_s)]\ge \exp \left\{-\frac{2\b}{\a(1-\a)} (\r(m)|\L|)^\a
[1-\l_-(\b)]\right\}.
\Eq(stimabassob)
$$

 Recalling \eqv(reduc), let us now give an upper bound for $\mu_\L^{++}[(\SS_\L(m,\e_0))^c|\SS^{VS}_\L(T, \r(m),\e_s)]$.
Recalling \eqv(Esse><),  we have
$$
 (\SS_\L(m,\e_0))^c={\cal M} ^>_\L(m,\e_0) \cup {\cal M} ^<_\L(m,\e_0).
 \Eq(partesse)
 $$
 We use again the Markov inequality to get on the one hand for $t_3>0$
 $$
 \mu_\L^{++}[{\cal M}^>_\L(m,\e_0)|\SS^{VS}_\L(T, \r(m),\e_s)]\le e^{-\b t_3(m+\e_0 m_\b)|\L|}
   \mu_\L^{++}[e^{+\b t_3 m_\L(\underline \s_\L)|\L|} |\SS^{VS}_\L(T, \r(m),\e_s)]
 \EQ(MVSup)
 $$
 and on the other hand, for $t_4>0$
 $$
 \mu_\L^{++}[{\cal M}^<_\L(m,\e_0)|\SS^{VS}_\L(T, \r(m),\e_s)]\le e^{+\b t_4(m-\e_0 m_\b)|\L|}
  \mu_\L^{++}[e^{-\b t_4 m_\L(\underline \s_\L)|\L|} |\SS^{VS}_\L(T, \r(m),\e_s)].
 \EQ(MVSdown)
 $$

Using Lemma \eqv(lpc) and Lemma \eqv(rho)  after some easy computations, if $|\L|$ is large
enough and how large depends on $\b,\a$, we have
 $$
 \mu_\L^{++}[\SS^>_\L(m,\e_0)|\SS^{VS}_\L(T, \r(m),\e_s)]\le e^{-\frac{\b}{8} \e_0 m_\b \z_\a |\L|^{\a+\g(1-\a)}}\equiv\eta_1(\b,\L)
 \EQ(MVSupt)
 $$
and by similar arguments,
we get
 $$
 \mu_\L^{++}[\SS^<_\L(m,\e_0)|\SS^{VS}_\L(T, \r(m),\e_s)]\le \eta_1(\b,\L).
 \EQ(MVSdownt)
 $$
Recalling \eqv(reduc), \eqv(stimabassob), \eqv(MVSupt) and \eqv(MVSdownt) we get \eqv(stimabasso). \eop

\bigskip

 To prove \eqv(main2), recalling \eqv(esseuno), let us define  the partition
 $$
 \left(\SS_\L^1(\r(m),\e_c)\right)^c=\SS_\L^<(\r(m),\e_c)\cup \SS_\L^>(\r(m),\e_c)
 \EQ(split)
 $$
 where
 $$
\SS_\L^<(\r(m),\e_c)= \bigcup_{\r< \r(m)-\e_c} \SS_\L^1(\r)
 \EQ(splitsm)
 $$
 and
 $$
 \SS_\L^>(\r(m),\e_c)= \bigcup_{\r> \r(m)+\e_c} \SS_\L^1(\r)
 \Eq(splitL)
 $$
 with the same conventions that are mentioned after \eqv(esseuno).

 \smallbreak

We consider first the set  \eqv(splitL), we have

 $$
 \mu^{+}_\L\left[\SS_\L^>(\r(m),\e_c),\SS_\L(m,\e_0)\right]\le \sum_{\r\ge \r(m)+\e_c} \mu^{+}_\L\left[S_\L^1(\r)\right]
 \EQ(avanti1)
 $$
 where the sum is merely over $\{\r\ge \r(m)+\e_c\}\cap Q_\L$.
  By similar computation as in the proof of Proposition \eqv(miming2) we have
 $$
 \mu^{+}_\L\left[S_\L^1(\r)\right]\le  \mu^{+}_\L\left[S_\L^B(\r)\right]
 [1+\eta_2(\b,\L)]
  \EQ(avanti2)
 $$
 where $S_\L^B(\r)$ is defined in \eqv(esseb),
$\eta_2(\b,\L)= \frac{ e^{-\b\frac{\z_\a}{4\a(1-\a)}(\e_c|\L|)^\a}
}{1-2e^{-\frac{\b}{\a(1-\a)} \e_0 c(\a) |\L|^\a}}
$
and   $c(\a)$ is defined in \eqv(calpha).
We have
  $$
  \mu^{+}_\L\left[S_\L^B(\r)\right]\le
\sum_{\underline T^E\in \TT^B_\L(\r)} e^{-\b h^{++}(\overline \s(\underline T^E))}\frac{\widetilde Z^{++}_\L(\underline T^E)}{Z^{++}_\L}
  \Eq(easy)
$$
where $\widetilde Z^{++}_\L(\underline T^E)$ is defined in \eqv(zeegamma).
Using cluster expansion, we get
$$
  \mu^{+}_\L\left[S_\L^B(\r)\right]\le
\sum_{\underline T^E\in \TT^B_\L(\r)} e^{-\b h^{++}(\overline \s(\underline T^E))[1-\l_+(\b)]}
  \Eq(easyb)
$$
where
$$
\l_+(\b)=2
\xi^{++} (\b) e^{4\b \zeta(2-\a)}
 (1+e^{-\frac{\b}{32} (\frac{\zeta_\a}{\a(1-\a)}-3\d)}).
\Eq(lpiu)
$$
and note that recalling \eqv(ksipiu), we have $\l_+(\b)\downarrow 0$ as $\b\uparrow \infty$
since $J$ defined in \eqv(2) is large and therefore we can assume that $J>\zeta(2-\a)$.
Using \eqv(reara3) we get that for all $\r \ge \r(m)+\e_c$, for $\underline T^E \in \TT^B_\L(\r)$,
$$
h^{++}(\overline \s(\underline T^E)\ge \frac{2}{\a(1-\a)} ((\r(m)+\e_c) |\L|)^\a
\Eq(Lreara4)
$$
and therefore using Lemma \eqv(entropy) and lemma \eqv(lb132), after a short computation,
if $|\L|$ is large enough and how large depends on $(m, \a, \b, \nu)$, we have
$$
\mu^{+}_\L\left[\SS_\L^>(\r(m),\e_c)|\SS_\L(m,\e_0)\right]\le e^{-\frac{2 \b (\r(m) |\L|)^\a}{\a(1-\a)} \left\{(1+\frac{\e_c}{\r(m)})^\a-1
\right\} }.
\Eq(last16b)
$$
where the $-1$ in the brackets comes from the lower bound \eqv(stimabasso) and we have assumed
$$
\a-\nu > \g
\EQ(condextra1234)
$$
to neglect the terms that come from Lemma \eqv(entropy) when performing the sum in \eqv(easyb).
This gives the second term in \eqv(main2).

\medskip
We consider now the set \eqv(splitsm) and  we write

$$
\eqalign{
 \mu^{+}_\L\left[\SS_\L^<(\r(m),\e_c),\SS_\L(m,\e_0)\right]&\le \mu^{+}_\L[\{m_\L(\underline \s_\L)\le m+\e_0m_\b\},\SS_\L^<(\r(m),\e_c)]\cr
&\le |\L| \sup_{\r\le \r(m)-\e_c}\left[ \mu^{+}_\L[\{m_\L(\underline \s_\L)\le m+\e_0m_\b\},\SS_\L^1(\r)] \right]\cr
&\le  |\L| \sup_{\r\le \r(m)-\e_c}\left[ \mu^{+}_\L[\{m_\L(\underline \s_\L)\le m+\e_0m_\b\}|\SS_\L^1(\r)] \right]\cr
}\EQ(start1)
$$
since $\mu^{+}_\L[\SS_\L^1(\r)] \le 1$.
Note that $\r(m) -\e_c>0$, by \eqv(epsmallzero) if $|\L|$ is large enough and how large depends only on $\b$ and $m$.

Using exponential Markov inequality, for $t>0$
we get
$$
\mu^{+}_\L[\{m_\L(\underline \s_\L)\le m+\e_0m_\b\}|\SS_\L^1(\r)] \le e^{t\b (m+\e_0m_\b)|\L|}
\mu^{+}_\L [e^{-\b t \sum_{i\in \L} \s_i}|\SS_\L^1(\r)].
\Eq(markless)
$$
To estimate the last term in \eqv(markless), we use the following Lemma.

\noindent{\bf \Lemma(titti)}
{\it There exists a $\b_{10} =\b_{10}(\a, \eta)$ such that for all $0<t\le  t^\star(\e_s)/2=
\frac{\z_\a }{4\a(1-\a) (\e_s |\L|)^{1-\a}}$, for all $\b\ge \b_{10}$,
{for all $\r<\r(m)-\e_c$}
 we have
$$
\mu^{+}_\L \left[e^{-\b t \sum_{i\in \L} \s_i}\big |\SS_\L^1(\r)\right]\le
\sup_{\underline T^E \in \TT^E_\L(\r)}
\mu_\L^+\left [e^{-\b t \sum_{i\in \L} \s_i}\big |\SS^{VS}_\L(\underline T^E,\r,\e_s )\right]
(1+ \eta_3(\b,\L))
\Eq(titti1)
$$
where $\TT^E_\L(\r)$ is defined in \eqv(calTrho), $\SS^{VS}_\L(\underline T^E,\r,\e_s )$ in \eqv(titti120pr)
 and
$$
\eta_3(\b,\L)= e^{-\frac{\b \z_\a}{8\a(1-\a)} (\e_s|\L|)^{\a}}= e^{-\frac{\b \z_\a}{8\a(1-\a)} (|\L|)^{\a(1-\g)}}.
\Eq(titti1b)$$
}

\medskip
Let us postpone the proof of this lemma and continue, using Lemma \eqv(lpc) and  Lemma \eqv(rho), inserting \eqv(titti1) in \eqv(markless),
using $\r\le \r(m)-\e_c$, $m(\r)=(1-2\r)m_\b$, and $m=m_\b(1-2\r(m))$, we get
$$
\mu^{+}_\L[\{m_\L(\underline \s_\l)\le m+\e_0m_\b\}|\SS_\L^1(\r)]\le
e^{-t\b (2\e_c -\e_0)m_\b|\L|}
(1+ \eta_3(\b,\L))
\Eq(titti12)
$$
 Taking $t= t^\star(\e_s) /2$, we get that the  exponential in the right hand side of \eqv(titti12) is
 $$
 e^{-\frac{\z_\a }{4\a(1-\a) (\e_s )^{1-\a}} \b(2\e_c-\e_0)m_\b|\L|^\a  }\le  e^{-\frac{\z_\a }{4\a(1-\a) } \b m_\b|\L|^{\a+\g(1-\a)-\nu }  }
\Eq(titti13)
$$
if $|\L|$ is large enough and how large depends only on $(a,\nu)$, see \eqv(epsmallzero).

\remark {\it We assume here $\e_c>\e_0$ that is $a>\nu$. }

Assuming
$$
\a+\g(1-\a)-\nu >\a,
\EQ(condextra123)
$$
using  \eqv(stimabasso) and recalling \eqv(start1),  if $|\L|$ is large enough, and how large depends on $(\a,\g,\nu)$, we get
$$
 \mu^{+}_\L\left[\SS_\L^<(\r(m),\e_c)|\SS_\L(m,\e_0)\right]\le
  e^{-\frac{\z_\a }{8\a(1-\a)} \b\left[ (2\e_c-\e_0)m_\b
  \right]
  |\L|^{\a+\g(1-\a)}  }
\Eq(titti2)
$$
where we have used a part of \eqv(titti13) to bound by 1 the terms coming from the factor
 $(1+\eta_3(\b,\L))$  and the lower bound \eqv(stimabasso).
This gives  the first term in \eqv(main2) .
\eop

\noindent{\bf Proof of Lemma \eqv(titti)}
\medskip
We start with
$$
\mu^{+}_\L [e^{-\b t \sum_{i\in \L} \s_i}|\SS_\L^1(\r)]
=\frac
{\sum_{\underline T^E \in \TT^E_\L(\r)} e^{-\b h^{++}(\overline \s(\underline T^E))}\sum_{\underline \s_\L\in \SS_{\underline T^E}}
e^{-\b \left[h^{++}(\underline \s_\L)-h^{++}(\overline \s(\underline T^E))\right]}e^{-\b t\sum_{i\in \L}\s_i}}
{\sum_{\underline T^E \in \TT^E_\L(\r) }e^{-\b h^{++}(\overline \s(\underline T^E))}\sum_{\underline \s_\L\in \SS_{\underline T^E} }
e^{-\b \left[h^{++}(\underline \s_\L)-h^{++}(\overline \s(\underline T^E))\right]}}
\Eq(titti117)
$$
where we used  \eqv(calTrho), \eqv(simplest), and  \eqv(essegae).

Now for each $\underline T^E \in \TT^E_\L(r) $,  see \eqv(calTrho), we have $\sum_{i\in \L} \overline \s_i(\underline T^E)=|\L|(1-2\r)$, because
$\underline T^E$ is made of mutually external triangles.
Therefore introducing
$$
\widetilde Z^{++}_\L(\underline T^E,-t)=  \sum_{\underline \s_\L\in {\SS}_{\underline T^E}}
e^{-\b \left[ h^{++}(\underline \s_\L)-h^{++}(\overline \s(\underline T^E)\right] }
e^{-\b t \sum_{i\in \L}\left[\s_i-\overline \s_i(\underline T^E)\right]}
\EQ(titti118)
$$
we get
$$
\eqalign{
\mu^{++}_\L [e^{-\b t \sum_{i\in \L} \s_i}|\SS_\L^1(\r)]&=e^{-\b t |\L| (1-2\r)} \times
\frac{\sum_{\underline T^E \in \TT^E_\L(\r) } e^{-\b h^{++}(\overline \s(\underline T^E))} \widetilde Z^{++}_\L(\underline T^E,-t)}
{\sum_{\underline T^E \in \TT^E_\L(\r) } e^{-\b h^{++}(\overline \s(\underline T^E))} \widetilde Z^{++}_\L(\underline T^E,0)}\cr
&\le e^{-\b t |\L| (1-2\r)} \times \sup_{\underline T^E  \in \TT^E_\L(\r) }
\frac{ \widetilde Z^{++}_\L(\underline T^E,-t)}{\widetilde Z^{++}_\L(\underline T^E,0)}
}\EQ(titti119)
$$
Therefore it remains to estimate the last ratio in \eqv(titti119) for $\underline T^E  \in \TT^E_\L(\r) $.
In a way similar to \eqv(verysmallT), let us define
$$
\SS^{VS}_\L(\underline T^E,\r,\e_s ) =\left\{\underline \s_\L \in \SS_{\underline T^E}\cap \SS_\L^1(\r) \,:\,
\forall \widetilde T \in \left\{ \underline T(\s_\L)\setminus \underline T^E \right\},
\,\, |\widetilde T|\le \e_s |\L|
\right\}
\EQ(titti120)
$$
and
$$
\widetilde Z^{++, VS}_\L(\underline  T^E,-t)=  \sum_{\underline \s_\L\in {\SS^{VS}_\L(\underline  T^E,\r,\e_s)}}
e^{-\b \left[ h^{++}(\underline \s_\L)-h^{++}(\overline \s(\underline T^E)\right] }
e^{-\b t \sum_{i\in \L}\left[\s_i-\overline \s_i(\underline T^E)\right]}
\EQ(titti121)
$$

We claim that, if $|\L|$ is large enough and how large depends on $\a,\b, \nu $ see \eqv(epsmallzero), we have
$$
\widetilde Z^{++}_\L(\underline T^E,-t)\le \widetilde Z^{++, VS}_\L(\underline T^E,-t) (1+\eta_3(\b,\L,\e_s))
\EQ(titti122)
$$
where
$\eta_3(\b,\L,\e_s)= e^{-
\frac{\b c_{10}(\a)}{C}
 (\e_s |\L|)^\a}$
with the choice $\e_s=|\L|^{-\g}$ we have $\eta_3(\b,\L,\e_s=|\L|^{-\g})=\eta_4(\b,\L)=e^{-
\frac{\b c_{10}(\a)}{C}
|\L|^{\a(1-\g)}}$.
Let us assume that \eqv(titti122) is true and continue.
Then the last ratio in \eqv(titti119) can be bounded as follows
$$
\frac{ \widetilde Z^{++}_\L(\underline T^E,-t)}{\widetilde Z^{++}_\L(\underline T^E,0)}\le (1+\eta_3(\b,\L,\e_s))
\times
\frac{\widetilde Z^{++, VS}_\L(\underline T^E,-t)}{\widetilde Z^{++, VS}_\L(\underline T^E,0)}
\Eq(titti122b)
$$
Recalling \eqv(titti119), we get immediately \eqv(titti1).  \eop

It remains to prove the claim  \eqv(titti122).
The proof is based  on an estimate of an energy cost when the magnetic field is negative and then a Peierls type argument.
Let us start with the energy estimate which is \eqv(titti114).
Recalling definition \eqv(SE=), given $0<\r\le 1$ and  $\underline T^E \in \TT^E_\L(\r) $
let us define
$$
\SS_{\underline T^E}(\r)=\big\{ \underline\s_\L\in \SS^1_\L(\r)  : \underline T^E(\underline \s_\L)=\underline T^E\big\}.
\EQ(titti3)
$$
As in section 6, for $\underline \s_\L \in\SS_{\G^E}(\r)$ there is a bijection between
$$
\underline \s_\L  \leftrightarrow (\underline T^E(\underline \s_\L), \underline \G(\underline \s_\L) )
\EQ(piti)
$$
where $\underline \G(\underline \s_\L)$ is the family of contours obtained by
implementing the algorithm ${\cal R} $ on $\underline T(\underline \s_\L) \setminus \underline T^E(\underline \s_\L) $.
Moreover,
$$
\underline \G(\underline \s_\L)=(\underline {\widetilde \G}(\underline \s_\L),\G^\star(\underline \s_\L))
\Eq(piti2)
$$
where $\underline {\widetilde \G}(\underline \s_\L) $ is a family of all contours made of  small  triangles {\it i.e.}
$\forall \G\in \underline {\widetilde \G}(\underline \s_\L) $, for all $ T\in \G$, $|T|\le \e_s|\L|$.
While, if they exist all the large triangles of $\underline T(\underline \s_\L) \setminus \underline T^E(\underline \s_\L) $, {\it i.e.}  the $T^\star$
such that $|T^\star| >\e_s|\L|$, are  internal with respect to $\underline T^E$ and belong to  the unique contour $\G^\star(\underline \s_\L)$.

The uniqueness of  $\G^\star(\underline \s_\L)$  is a consequence of \eqv(4.1b). In fact,
picking up  two of them, say
$\G^\star_1, \G^\star_2$, using \eqv(4.1b) and the fact that $0<\g <2/3$  we should have
$$
{\rm dist}(\G^\star_1, \G^\star _2)> C {\min} \{ |\G^\star_1|^3, |\G^\star_2|^3\} \ge  C  \e_s^3|\L|^3>|\L|
\EQ(onlyone)
$$
which is not possible for two contours having their supports within $\L$.

The fluctuations with respect to the ground state $\overline \s(\underline T^E)$ are  therefore described by the
family of contours
$(\underline \G(\underline \s_\L), \G^\star(\underline\s_\L))$.
Using  the bijection between spin configurations and contours we write
$$
h^{++}(\underline \s_\L)-h^{++}(\overline \s(\G^E)+ t\sum_{i\in \L}\s_i
=H(\underline \G, \G^\star)+ t\sum_{i\in \L} \s_i(\underline \G, \G^\star)
\EQ(titti111)
$$

Consider now the basis of the triangles belonging to $\G^*$, they define a set of intervals whose interior
boundary is made of spins with the same sign.
Calling $I^+$ (resp. $I^-$) the union of intervals with interior boundary $+$ (resp $-$) we have that $\D(\G^\star)=I^+\cup I^-$.
If we define the transformation $\t^{\star}$ that depends on $\G^\star$ by :
If $i\in \D(\G^\star)$
$$
\t^{\star}_i (\underline \s_\L)=\cases{-\s_i, &if $i\in I^+$;\cr
\s_i, & otherwise.\cr}
\EQ(titti112)
$$
while if $i\in \L\setminus \D(\G^\star)$ then $\t^{\star}_i (\underline \s_\L)=\s_i$, then we get
$$
h^{++}(\t^{\star} (\underline \s_\L))-h^{++}(\overline \s(\G^E))+t\sum_{i\in \L}\t^{\star}_i (\underline \s_\L)
=
H(\underline \G)+t\sum_{i\in \L}\t^\star_i(\underline \G)
\EQ(titti113)
$$
so that the excess of energy when the magnetic field is negative,  due to the presence of $\G^\star$ is
$$
H(\underline \G, \G^\star)-H(\underline \G) +
\,t  \sum_{i\in I^+} [\s_i(\underline \G,  \G^\star)-\t^\star_i(\underline \G)]
\ge
\frac{c_{10}(\a)}{C}
\|\G^\star\|_\a + 2t \sum_{i\in I^+} \s_i(\underline \G,  \G^\star)
\ge
\frac{c_{10}(\a)}{C}
\|\G^\star\|_\a
\EQ(titti114)
$$
where we have used \eqv(MP333),  $c_{10}(\a) =\frac{2\pi^2}{3\a(1-\a)}$ and the fact that
$
\sum_{i\in I^+} \s_i(\underline \G, \G^\star)
$ is positive.

To check this last fact, we remark that a necessary condition for a spin $\s_i(\underline \G, \G^\star)$
at $i \in I^+$ to be negative is that it belongs to a contour
of $\underline \G$.
Since the number of contours of mass $m$, see \eqv(mass),  that are in $I^+$  is less or equal to $|I^+|/(C m^3)$
because the inter-distance between two such contours is larger than $C m^3$, see \eqv(4.1b).
On the other hand, it is easy to see that
the number of negative spins in a contour is smaller or equal to  the mass of this contour with equality
for contours made of mutually external triangles.
Therefore if we denote by
$M^-(I^+)(\underline \s)=\sum_{i \in I^+} \frac{1-\s_i}{2}$
and $M^+(I^+)(\underline \s)=\sum_{i \in I^+} \frac{1+\s_i}{2}$
 the number of negative,  respectively positive spins in $I^+$,
we have
$$
M^-(I^+)(\underline \s(\underline \G,  \G^\star))
\le
\frac{|I^+|}{C} \sum_{m=1}^\infty \frac{m}{m^3}= \frac{ \pi^2|I^+|}{6 C}.
\EQ(titti115)
$$
Since
$M^-(I^+)(\underline \s(\underline \G, \G^\star))
+M^+(I^+)(\underline \s(\underline \G,  \G^\star))=|I^+|$,  from \eqv(titti115) we get
$$
\sum_{i\in I^+}\s_i(\underline \G, \G^\star)
=|I^+|-2M^-(I^+)(\underline \s(\underline \G,  \G^\star))\ge |I^+| \left(1-\frac{\pi^2}{3C}\right)>0
\Eq(titti116)
$$
if $C >\pi^2/3$, that we can assume. This ends the proof of \eqv(titti114).

\medskip
The above mentioned Peierls type argument runs as follows:
Let us write
$$
\widetilde Z^{++}_\L(\underline T^E,-t)=\widetilde Z^{++, G}_\L(\underline T^E,-t)+ \widetilde Z^{++, VS}_\L(\underline T^E,-t)
\EQ(titti123)
$$
where,
$$
\widetilde Z^{++, G}_\L(\underline T^E,-t)=
\sum_{\underline \s_\L\in {\SS^{G}_{\underline T^E}}}e^{-\b \left[ h^{++}(\underline \s_\L)-h^{++}(\overline \s(\underline T^E)\right] }
e^{-\b t \sum_{i\in \L}\left[\s_i-\overline \s_i(\underline T^E)\right]}
\EQ(titti124)
$$
and
$$
\SS^{G}_{\underline T^E}=\SS_{\underline T^E}(\r) \setminus \SS^{VS}_{\L} (\underline T^E, \r, \e_s),
\EQ(titti125)
$$
see \eqv(titti120) and \eqv(titti3).
We have
$$
\widetilde Z^{++, G}_\L(\underline T^E,-t) \le \sum_{\G^*\sim \underline T^E}\,\,
\sum_{\underline {\widetilde \G} \sim \G^*\cup\underline T^E
}
e^{-\beta H(\G^*,\underline {\widetilde \G}) +t\sum_i \s_i(\G^*, \underline {\widetilde \G})}
\Eq(tipi1234)
$$
where  the first sum is over the $\G^*\sim \underline T^E$ that is the set of $\G^\star$ such that
there exists a configuration $\underline \s_\L$ such that, recalling \eqv(piti2)
$$
\G^\star(\underline \s_\L)=\G^\star \,\,{\rm and}  \,\, \underline T^E(\underline \s_\L)=\underline T^E
\EQ(compati)
$$
and analogous definition for the second sum.
Using  \eqv(titti114)
$$
\widetilde Z^{++, G}_\L(\underline T^E,-t)\le
\sum_{\G^*\sim \underline T^E}e^{-\beta
\frac{c_{10}(\a)}{C}
\|\G^*\|_\a}
\sum_{\underline {\widetilde \G} \sim \G^*\cup\underline T^E
}
e^{-\beta H(\underline {\widetilde \G}) +t\sum_i \s_i( \underline {\widetilde \G})}
\Eq(tipi1235)
$$
and therefore using \eqv(eq:th2.3)
$$
\eqalign{
\widetilde Z^{++, G}_\L(\underline T^E,-t)&\le
\sum_{\G^*\sim \underline T^E}e^{-\beta
\frac{c_{10}(\a)}{C}
\|\G^*\|_\a}Z^{++,VS}(\underline T^E,-t)\cr
&\le 2 |\L|
e^{-\beta
\frac{c_{10}(\a)}{C}
(\e_s |\L|)^\a }Z^{++,VS}(\underline T^E,-t).\cr
} \EQ(tipi1236)
$$
which is \eqv(titti122). \eop


 \smallbreak
 \chap { 6 Appendix 1: Triangles, contours and Polymers}6
\numsec= 6 \numfor= 1
\numtheo=1

In this section we regroup all the definitions and estimates that comes from [\rcite{CFMP}] for the Peierls argument
and from [\rcite{CMPR}] for the cluster expansion.

\noindent {\bf 6.1 Triangles configurations}

 In this section we start recalling the content of  [\rcite{CFMP}].
For all $i^*\in \L^*$, we consider  an interval $[i^*- \frac 1 {100}, i^* + \frac 1 {100} ] \subset \R$
and choose one  point in each interval, say $r_{i^*}\in \R$  in such a way that
for any four distinct points $r_j$,  $j=1, \dots, 4$ $|r_1-r_2| \neq |r_3-r_4|$.

  Given a  spin configuration $\underline \s_\L \in \SS_\L$, consider the set of its spin flip points $\LL^*(\s_\L)$
  {\it i.e.} the set of $i^* \in \L^*$ such that $\s_{i^*-\frac{1}{2}}=-\s_{i^*+\frac{1}{2}}$
  and the corresponding points $(r_{i^*}, i^* \in \LL^*(\s_\L))\subset \R$.

   We next embed  $\R$ in $\R^2$
 where the line containing the $r_{i^*}$  represents the state at $t=0$, and
 the orthogonal axis represents the evolving
 time of a  process of growing ``$\vee$-lines":   each point $r_{i^*}$
  branches into two twin  lines growing at velocity $1$ in the positive half plane, in the directions
 respectively of angles $\pi/4$ and $3/4 \pi$, until one of the
 two meets another line coming from a different $r_{j^*}$.
 At the instant  when two branches of different $\vee$-lines meet,
 they are frozen and stop their growth, at the same instant their twin lines
disappear, while all the other $\vee$-line associated to the other  points
are undisturbed and keep growing.
The collision of two lines  is represented graphically in the $(r,t)$
plane by a triangle
whose basis is the interval between the two points $r_{i^*},r_{j^*}$,
roots of  the two lines that  collided,
 and the third vertex  is the point representing the
collision  in the plane $(r,t)$.

This construction is a way to construct  a pairing of  spin flips
with a  criterion of minimal distance.
Our choice of   $r_{i^*}$ makes the definition of triangles non--ambiguous.

For any finite $\L$
the process stops   at a finite time $t\le |\L|+1$ giving  rise to  a configuration  of  triangles.
The  triangles will be denoted by $T$ and a family of triangles will be denoted by $\underline T$.

\smallbreak
\Definition(triangles) {$\,$} {\it Given a spin configuration $\underline \s_\L \in \SS_\L$, we denote by $\underline T(\underline \s_\L)$
the configuration of triangles obtained following the above mentioned procedure. }
\smallbreak

$x_-, x_+$ will denote
respectively the left and right root of the associated  $\vee$-lines.
  We will also write:

 $$
 \Delta(T)=[x_-(T),x_+(T)] \cap \Z \quad   \hbox{the basis of the triangle}\,\, T;
 \Eq(def:delta-T)
 $$
 $$
 |T|=\#\{\Delta(T)\}\equiv |\Delta(T)|  \quad   \hbox {the mass of the triangle}\,\, T;
 \Eq(def:|T|)
$$
$$
{\rm sf}(T)= \big\{
\inf( \Delta(T))-1, \inf( \Delta(T)),
\sup( \Delta(T)), \sup( \Delta(T))+1\big\}
\Eq(def:sspT-a)
$$
where $\Z $ is equipped with its  natural order;
$$
{\rm dist}(T,T')= {\rm dist} ({\rm sf}(T),{\rm sf}(T')).
\Eq(def:dist-T)
$$
From our construction it follows that for all triangles $T_i\ne T_j$,
 $$
{ \rm dist}(T_i,T_j)\ge\min(|T_i|, |T_j|).
\Eq(prop:dist-T)
 $$

In particular if a triangle $\widetilde T$ is interior to a triangle $T$, {\it i.e.} is such that $\D(\widetilde T)\subset \D(T)$, then
$$
|\widetilde T|\le \frac13 |T|.
\EQ(unterzo6)
$$

We denote $\TT_\L $ the set of configurations of
 triangles $\underline T =(T_1,\dots, T_n)$ that satisfy  \eqv(prop:dist-T) and such that  $\D(T_i)\subset \L $
 for all $i\in \{1,\dots,n\}$.
Since here the spins at the boundary are specified, $\bar \s_{-L-1}=\bar \s_{L+1} =1$,
where we recall that $\L=[-L,+L]\cap \Z$,
the above construction defines a one to one map from $ \SS_\L$ to $ \TT_\L$.
In particular, if $\1$ denotes the spin configuration in $\L$ constantly equals to $+1$,
 $\1$ is mapped to the empty configuration of triangles.

 We say that  two collections of triangles $\underline S'\in   {\cal T}_\L $  and $\underline S\in   {\cal T}_\L $
  are
 compatible  and we denote it by  $\underline S' \simeq \underline S$  iff
$ \underline S' \cup \underline S \in  {\cal T}_\L $
({\it i.e.}
there exists a configuration in $\SS_\L$ such that its corresponding collection  of triangles
is the collection made of all triangles  in $\underline S'$ and   $\underline S$.)

The basic estimates for a collection of triangles $\underline T\in \TT_\L$ is given in Lemma 2.1
and appendix A.1 of [\rcite{CFMP}]:
calling
$$
H^{++}(\underline T)=h^{++}(\s_\L(\underline T)
\Eq(renorm)
$$
note that from \eqv(1),  the configuration with no triangles has an energy $0$ and is a ground state.
\smallbreak

\noindent{\bf \Lemma(tr1)}
{\it For all $0<\a\le -1+(\log 3/\log 2)$, for $J$ large enough, for all $\underline T\in \TT_\L$
$$
H^{++}(\underline T)\ge \frac{2 \z_\a}{\a(1-\a)} \sum_{T\in \underline T} |T|^{\a}
\Eq(lbt)
$$
where $\z_\a=1-2(2^\a-1)>0$.

}

The proof is given in Lemma 2.1 and appendix A.1 of [\rcite{CFMP}]
by exploiting the property \eqv(prop:dist-T). Note that in [\rcite{CFMP}]  formula (3.4),  there is a  misprint
 where $\le$ should be replaced by $\ge$. Note also the factor $2/\a(1-\a)$ that was present in
appendix A in [\rcite{CFMP}], see the proof of Lemma A.1 there, is missing in formulae (3.4) and (2.9) in [\rcite{CFMP}].
Unfortunately this missing factor $\a(1-\a)$ propagate also in [\rcite{CMPR}].

Let us now give the

\noindent {\bf Proof of Lemma \eqv(entropy)}
Let us make a partition of the interval $\L$ in segments  of size $\e_s|\L|/2$, there are less than $2|\L|/(\e_s |\L|)$ such segments.
Given a family of triangles $\underline T^E\in \TT^E_\L(\r)$, see \eqv(calTrho), let $n(\underline T^E)$ be the maximum number of
segments contained in $\D(\underline T^E)=\cup_{\tilde T \in \underline T^E} \D(\tilde T)$.
We have $ \r|\L|/(\e_s|\L|)\le n(\underline T^E)\le 2\r|\L|/(\e_s|\L|)$.
Given an integer $n$ and a family of $n$ such segments, the number of families of external  triangles that can contain this particular
family of segments is less than $(\e_s|\L|)^n$. Then taking $\e_s=|\L|^{-\g}$ we get
$$
 \sharp  [\TT^E_\L(\r)]\le
\sum_{n=  \r|\L|/(\e_s|\L|)}^{2 \r|\L|/(\e_s|\L|)} {|\L|^\g \choose n} (\e_s|\L|)^n\le
|\L|^\g (|\L|^{1-\g})^{2\r |\L|^\g} 2^{|\L|^\g}
\le  e^{(2-\g) |\L|^\g \log |\L|}
\EQ(entro2)
$$
if $|\L|$ is large enough and how large depends only on $\g$.
\eop
\smallbreak

 A contour will be  a family of triangles $\G\equiv \{T: T\in \G\}$ that satisfy the  properties
listed in the Definition \eqv(cont).

Let us define
$$
|\G|\equiv \sum_{T\in \G}|T|\quad \hbox{the mass of the contour}
\Eq(mass)
$$
and for $\a>0$ we define
$$
\|\G\|_\a\equiv \sum_{T\in \G}|T|^\a.
\Eq(massalpha)
$$
Recalling  \eqv(def:delta-T) to \eqv(def:dist-T),   let us denote
$$
\Delta(\G)\equiv \bigcup_{{T\in \G}}\Delta(T)
\Eq(def:delta-Ga)
$$
$$
x_{-}(\G)\equiv\min_{T\in \G}x_{-}(T)
\Eq(def:x--Ga)
$$
$$
{\rm sf}(\G)\equiv \bigcup_{T\in \G}{\rm sf}(T)
\Eq(def:ssp-Ga)
$$
$$
{\rm dist}(\G,\G')\equiv \inf_{{T\in\G}\atop {T'\in\G'}}{\rm dist}(T,T')
\Eq(def:dist-Ga)
$$
$$
\underline  T(\G)=\{T: T\in \G\}
\Eq(def:T-Gab)
$$

\smallbreak
\noindent{\bf \Definition(cont).}
{\it Given a configuration of triangles $\underline T$ in $\TT_\L$, a configuration of contours
$\G=\G(\underline T)$ is the result of the implementation of an algorithm $\RR$ on the family of triangles $\underline T$, denoted by
$\underline \G(\underline T) =\RR (\underline T)$.
It is a partition of $\underline T$ whose atoms, called {\rm contours} are
determined by the following
properties  $P.0, P.1, P.2$ :
\smallbreak
\noindent{\bf P.0} {\it  Let ${\RR}(\underline T) \equiv
( \G_1,..,\G_n)$, $\G_i =\{ T_{j,i}, 1\le j
 \le k_i\}$, then $\underline T=\{T_{j,i}, 1\le i \le n,\;1\le j
 \le k_i\}$}.
\smallskip
\noindent {\bf P.1} {\it Contours are well separated from each
other.} Any pair $\G\ne\G'$ in $\RR(\underline T)$
verifies one of the following two alternatives. (i):
$\Delta(\G)\cap \Delta(\G')=\emptyset$, or (ii):
$\Delta(\G)\cap \Delta(\G')\ne \emptyset$, then either
$T(\G)\subset \Delta(\G')$ or
 $T(\G')\subset \Delta(\G)$;
 moreover, supposing for instance that the former case is verified,
 (in which case we call $\G$ an {\rm inner} contour, $\G'$ is {\rm  external} w.r.t $\G$),
 then  for any triangle $T'_i\in \G'$, either
$T(\G)\subset T'_i$ or $T(\G)\cap T'_i=\emptyset$.
Namely either $\Delta(\G)\cap \Delta(\G')=\emptyset$ or
$$
\sum_{T'\in \G'}\1_{\{T(\G)\subset \Delta(T')\}}+\sum_{T\in \G}\1_{\{T(\G')\subset \Delta(T)\}}=
1.
\Eq(pe1)
$$
In both cases
$$
{\rm dist}(\G,\G') > C \,\min\left\{|\G|^3,|\G'|^3\right\}
\Eq(4.1b)
$$
where  $C$  is a constant chosen such that,
as  in { \rm [\rcite{CFMP}]}, {\it we have
$$
\sum_{M} \frac{4M}{[C M^3]}\le \frac 12
\Eq(3.9)
$$
and  dist$(\G,\G')$ is defined in \eqv(def:dist-Ga).}
\smallskip
\noindent {\bf P.2} {\it  Independence.} Let $\{\underline T^{(1)},\dots,\underline
T^{(k)}\}$, be $k> 1$ configurations of triangles;
$\RR(\underline T^{(i)})=\{\G^{(i)}_j, j=1,\dots,n_i\}$ the
contours of the configuration $\underline T^{(i)}$.  Then, if any
distinct pair $\G^{(i)}_j$ and $\G^{(i')}_{j'}$ satisfies
$P.1$
$$
\RR\left (
\underline T^{(1)},\underline T^{(2)},\dots,\underline T^{(k)}
\right)=\{\G^{(i)}_j, j=1,..,n_i; i=1,..,k\}.
\Eq(4.2e)
$$
}

\noindent {\bf \Definition(compga)}.
  (compatibility between contours) $\,\,${\it We say that two contours $\G, \G'$ are compatible if \eqv(4.1b) is verified. We denote
$$
\eqalign{
\G \sim \G' \quad & \Longleftrightarrow\quad
      \G,\G' \hbox{ are compatible}\cr
\G \not\sim \G' \quad &\Longleftrightarrow\quad
      \G,\G' \hbox  {are incompatible}.\cr
}
\Eq(compaga)
$$
}

Therefore we have a bijection between spin configurations in $\SS_\L $ and triangles in $\TT_\L$
and another one between $\TT_\L $ and its image by $\RR$, in particular there is a one--to--one
correspondence between spin configurations
 and contour configurations. We denote by $\GG_{\L}$  the set of all possible configurations of compatible contours
associated to $\SS_{\L}$.

 Moreover
denoting by $\underline T(\underline \G)$ the configuration  of triangles that  are in $\underline \G$,
we define
$$
H^{++} (\underline \G)\equiv H^{++} (\underline T(\underline \G))
\Eq(tuttialmare)
$$

 For $\underline \G \in \GG_{\L}$ let
$
\underline \s_\L(\underline \G)\,\,
 $
be  the  corresponding  spin configuration.

\smallbreak
\Definition(def:ext)  $\,\,$ {\it Given a collection of contours $\underline \G$, a contour $\G\in \underline \G$ is external
with respect to
 $\underline \G$ if  each  triangle of $\D(\G)$ is  not contained in some triangle
that belongs to the others contours of $\underline \G$.
  $$
  \forall T \in \G ,   \quad \D(T) \not\subset \bigcup_{\G'\in \underline \G :  \G'\ne \G} \Delta(\G')
  \Eq(eq:def-ext)
$$
A contour which is not external is called {\it internal}.}

The following Proposition regroups all the basic estimates that we
need.
It is proved in [\rcite{CFMP}].

\smallbreak
 \noindent{\bf \Proposition (thm:2.5)}
$\,${\it
If $0\le \a<\a_+=-1+(\log 3)/(\log 2)$, then
for any contour $\G $,
$$
H^{++}(\G)
\geq
\frac{\z_\a}{\a(1-\a)}
\|\G\|_\a
\Eq(eq:T2.5-1)
$$
with $\zeta_\a$ as in Lemma \eqv(tr1).
and $\|\G\|_\a$ is defined by \eqv(massalpha).

\smallskip
For any familly of compatible contours $(\G_0,\underline \G)$,
$$
H^{++}(\G_0\cup\underline \G)-H^{++}(\underline \G)\ge \d \|\G_0\|_\a
\EQ(MP333)
$$
where if $0<\a<\a_+ $,
$\delta \equiv c_{10}(\a) /C$ with $C$ as in \eqv(4.1b) and $c_{10}(\a)=2\pi^2(3\a(1-\a))^{-1}$

There exists a $b_0\equiv b_0(\a)$ such that for all $b\ge b_0$, for all integers $m\ge 1$,
$$
\sum_{{\scriptstyle \G : x_-(\G)=0}\atop{\scriptstyle |\G|=m}}e^{-b \|\G\|_\a} \le 2 e^{-b m^\a}.
\Eq(eq:th2.3)
$$}

\noindent {\bf Cluster expansion}

Let us recall the result on the cluster expansion given  in [\rcite{CMPR}].
Section 4 of [\rcite{CMPR}] gives the derivation of the partition function as that of a gas of polymers with
an hard core condition and Proposition 5.4 there proves that the cluster expansion for the logarithm of the partition function
is convergent is
$$
\sum_{ R\ni 0}\xi^{++}( R) <<1
\Eq(onepoli)
$$
where $\xi^{++}( R)$ is the activity of the polymer $R$ and the sum is over al the polymers containing the origin.

In [\rcite{CMPR}] it is proved that \eqv(onepoli)
follows basically  from \eqv(lbt), or modification of it as  \eqv(eq:T2.5-1) by  taking $\beta$ large enough.
The consequence of the convergence of the cluster expansion is that we have
 $$
\log Z_{\L}^{++}=
\sum_{x\in  \L}
\xi^{++}(R_x^1) \left[1+{ \cal B}(x,++)\right]
\EQ(Mp61c)
$$
where:

1)
$R_x^1$ is the triangle of size $1$ located in $x$, i.e.  is the simplest contour  and also the simplest polymer made of a singleton
and
$$\xi^{++}(R_x^1)=
e^{-2\b(J+\z(2-\a))}.
\Eq(MC5)
$$
where $\z(2-\a)$ is the Riemann zeta function and $J$ is defined in \eqv(2).

2) ${ \cal B}(x,++)$ is an absolutely convergent series
 $$
 { |\cal B}(x,++)| \le e^{-\frac{\b}{32}(\frac{\z_\a}{\a(1-\a)}-3\d)}
 \Eq(est1triangle)
 $$
 where $\z_\a$ is the same as in lemma \eqv(tr1) and $\d$ is the same as in \eqv(MP333).

 \smallbreak
 \smallbreak
Here we are considering instead of \eqv(Mp61c) the logarithm of a constrained partition function as

$$
{\widetilde Z^{++}_\L(\underline T^E,t)}=  \sum_{\underline \s_\L\in {\SS}_{\underline T^E}}
e^{-\b \left[ h^{++}(\underline \s_\L)-h^{++}(\overline \s(\underline T^E)\right] }
e^{\b t \sum_{i\in \L}(\s_i-\overline \s_i(\underline T^E))}
\EQ(laplace32)
$$

the partition function restricted to the set of configurations ${\SS}_{\underline T^E}$, see \eqv(essegae),  where an external field
$t$ is present. Here $\underline T^E\in {\cal T}^E_\L(\r)$, see  \eqv(calTrho) for some $0<\r<1$.

In the case $t=0$, the ground state is   $\bar\s(\underline T^E)$, see \eqv(simplest) and \eqv(GS)  and the energy fluctuations
 $\underline T$ are in
 $$
 {\cal T}_\L^f(\underline T^E)=\left\{ \underline {\widetilde T} \in {\cal T}_\L : \underline {\widetilde T} \sim \underline T^E,
 \underline T^E[\underline \s_\L ( \underline{\widetilde  T} \cup \underline T^E)]=\underline T^E
 \right\}
 \Eq(fluctuat)
 $$
 that is the set of family the triangles $\underline {\widetilde T}$ that are compatible with $ \underline T^E$, their presences does not modify the
 family of external triangles $\underline T^E$ in particular the triangles of  $\underline {\widetilde T}$ that are external to $\underline T^E$  are all small.

  Let us define, for $\underline T \in  {\cal T}_\L^f(\underline T^E)$
   $$
  H^{++}_{\underline T^E} (\underline T)=h^{++} (\underline \s(\underline T \cup \underline T^E))-h^{++}(\overline \s(\underline T^E))
  \EQ(excess)
  $$
 that will be simply denoted $  H^{++} (\underline T)$ if no confusion could arise.
 One can check  that if $\underline \s_\L \in S_{\underline T^E}$ and $\underline {T}=\underline {T}(\underline \s_\L)$ is
  the associated family of triangles in ${\cal T}_\L^f(\underline T^E)$
 then
  $$
h^{++}(\underline \s_\L)-h^{++}(\overline \s(\underline T^E)-t \sum_{i\in \L}(\s_i-\overline \s_i(\underline T^E))
\ge
H^{++}(\underline T) -|t|\sum_{T \in \underline T} | T|.
\EQ(onesteptoheaven)
$$

Therefore, in presence of a magnetic field $t$ as in \eqv(laplace32), an analogous of the condition \eqv(lbt) is:
For all $\underline T\in \TT^f_\L(\underline T^E)$
$$
H^{++}(\underline T)-|t|\sum_{T\in \underline T}| T|\ge \frac{ \z_\a}{\a(1-\a)}
\sum_{T\in \underline T} |T|^{\a},
\Eq(lbt-E)
$$
This condition is satisfied if
for all $\underline T\in \TT^f_\L(\underline T^E)$
$$
\frac{2 \z_\a}{\a(1-\a)}  |T|^{\a}-|t||T|\ge \frac{ \z_\a}{\a(1-\a)} |T|^{\a},
\Eq(lbt-E2)
$$
i.e. if
$$
|t|\le
\frac{ \z_\a}{\a(1-\a)}  (\frac13 |\underline T^E|)^{\a-1}\le
\inf_{T\in {\cal T}^f_\L(\underline T^E)}
\frac{ \z_\a}{\a(1-\a)}  |T|^{\a-1}.
\Eq(lbt-E3)
$$
where we have used \eqv(unterzo).

We regroup in the following Lemma the contribution of the dominant term of the constrained free energy obtained by cluster expansion.
The proof is the same as the one of \eqv(Mp61c) given in [\rcite{CMPR}]. On the other hand since all polymers that can occur in the expansion of
$\log \widetilde Z^{++}_\L(\underline T^E,t )$ occur also in the expansion of \eqv(Mp61c), the error terms satisfy the same bound.

 \smallbreak
\noindent{\bf \Lemma(Lfond)}
{\it There exists a $\b_1=\b_1(\a) $ such that if $\b\ge \b_1(\a)$,
 for all $\r\in [0,1]\cap Q_\L$, for $\e_ s=|\L|^{-\g}$ with $0<\g<2/3$ and for $\L$ so large to have $\r>\e_s$  then,
 for
 $t\le \z_\a 3^{1-\a} /(4\a(1-\a) (\r |\L|)^{1-\a})\equiv t^\star_{\L,1}(\r)$
 where $\z_\a=1-2(2^\a-1)>0$  we have, for
 all $\underline T^E\in {\cal T}^E_\L(\r)$,
  $$
\log \widetilde Z^{++}_\L(\underline T^E,t )=\sum_{x\in \L} \1_{\{{\rm dist}(x,sf(\underline T^E))\ge 1\}} \xi^{\overline \s(\underline T^E)}(x) e^{-2\b t \overline \s_x(\underline T^E) } (1+{\cal B}(x, \underline T^E, t))
\Eq(CEt2b)
$$
where $\xi^{\overline \s(\underline T^E)}=e^{-\b [h^{++}(T^{\{x\}}\overline \s(\underline T^E))-h^{++} (\overline \s(\underline T^E))]}$ and $T^{\{x\}}\overline \s(\underline T^E)$ is the configuration
equals to $\overline \s(\underline T^E)$ everywhere
but at $x$ where   the spin $\overline \s_x(\underline T^E)$ is reversed and
 ${\cal B}(x, \underline T^E, t)$ can be written as an explicit absolutely convergent series and satisfies
$$
|{\cal B}(x, \underline T^E, t)| \le e^{-\frac{\b}{64}(\frac{\z_\a}{\a(1-\a)}-3\d)}
\EQ(bbxitzerob)
$$
where as  in Proposition \eqv(thm:2.5),
$$
\d=2\pi^2(3\a(1-\a))^{-1} C^{-1}
\Eq(delta007)
$$
and $C$ is the constant that appears in the definition of contours
\eqv(4.1b)}.
\vskip .5cm
{\bf Remark:} For future reference we notice that: if $x\in \L$

$$
\xi^{\overline \s(\underline T^E)}(x) =\xi^{++}(\b) \times
\cases{e^{2\b \sum_{y\in \Z \setminus \D(\underline T^E)} J(x-y)}, & if  $x\in \D(\underline T^E)\setminus {\rm sf}(\underline T^E)$;\cr
e^{2\b \sum_{y\in \D(\underline T^E)\setminus \{x\} } J(x-y)}, & if  $x\notin \D(\underline T^E) \cup  {\rm sf}(\underline T^E)$.\cr
}
\Eq(xitzerob)
$$

\vskip 1cm

Let us now prove Lemmata \eqv(rho) and \eqv(lpc)

For the proof of \eqv(fondarhobis),
given $\underline t=(t_i, i\in \L)\in \R^\L$, let us define  for $\underline T^E \in \TT^E_\L(\r)$,
$$
{\widetilde Z^{++}_\L(\underline T^E,\underline t)}=
\sum_{\underline \s_\L\in {\SS}_{\underline T^E}}e^{-\b \left[ h^{++}(\underline \s_\L)-h^{++}(\overline \s(\underline T^E)\right] }
e^{\b \sum_{i\in \L}t_i(\s_i-\overline \s_i(\underline T^E))}.
\EQ(laplace32-b)
$$

By elementary computations, one can check that
$$
\mu_{\L}^{++}[\s_i|\SS_{\underline T^E}]=\overline \s_i(\underline T^E) + \frac{1}{\b}
\frac{\partial }{\partial t_i}[\ln {\widetilde Z^{++}_\L(\underline T^E,\underline t)}]\bigg|_{\underline t=0}.
\Eq(cumu)
$$
Then it is immediate to check that
$$
\frac{1}{|\L|} \sum_{i\in \L} \overline \s_i(\underline T^E)=(1-2\r)
\EQ(const)
$$
since $\overline \s(\underline T^E)$ is made of  $\r|\L|$ sites in $\L$ with $-1$ and what remains is $+1$.
Now recalling \eqv(CEt2b),  the dominant terms that comes from the second term in \eqv(cumu) gives a contribution to the mean
$$
\frac{1}{|\L|} \sum_{i\in \L} (-2\overline \s_x(\underline T^E))  \1_{\{{\rm dist}(x,sf(\underline T^E))\ge 1\}} \xi^{\overline \s(\underline T^E)}(x).
\Eq(MMP)
$$
Recalling \eqv(ksipiu) and writing
$\xi^{\overline \s(\underline T^E)}(x) =\left[\xi^{\overline \s(\underline T^E)}(x) -\xi^{++}(\b)\right]  + \xi^{++}(\b)$ in \eqv(MMP),
to the term with $\xi^{++}(\b)$ corresponds
$$
\frac{1}{|\L|} \sum_{i\in \L} (-2\overline \s_x(\underline T^E))\xi^{++}(\b)=
-2\xi^{++}(\b)(1-2\r).
\EQ(MMP2)
$$
Recalling \eqv(preliminary), we get \eqv(fondarhobis) since, by similar arguments as in the end of section 4
$$
\frac{2}{|\L|} \sum_{i\in \L} \left | \xi^{\overline \s(\underline T^E)}(x) -\xi^{++}(\b)\right| \le \frac{10\xi^{++}(\b) }{\a(1-\a)} |\L|^{\a-1}.
\Eq(grandfinal)
$$
from which we get \eqv(fondarhobis). Using \eqv(hatrho) we get \eqv(fondarho).

Let us now complete the proof the Lemma 3.2.
It is enough to have an estimate uniform in the magnetic field $r$ which satisfies \eqv(largestt) or \eqv(smallestt) for the two points truncated correlation function \eqv(covarb). For notational simplicity,  we consider only the case of \eqv(lpc1), the case of \eqv(lpc1a) can be proved {\it mutatis mutandis}.
Here we have to start with a $T_0$ with $|T_0|=\r|\L|$.
Considering ${\widetilde Z^{++}_\L(T_0,\underline t)}$ as in \eqv(laplace32-b)
we take first $ \underline t=r+\underline t(i,j)$ where
$t_k(i,j)=0$, if $k\neq i, j$ while $t_i(i,j)=t_i$, $t_j(i,j)=t_j$ and then we get easily
$$
\mu^{+}_\L(r)[\s_i,\s_j| \SS_{T_0}]= \frac{1}{\b^2} \frac{\partial}{\partial t_i \partial t_j} \log  {\widetilde Z^{++}_\L(T_0,r+t(i,j))}
\bigg|_{t(i,j)=0}
\EQ(pets1)
$$
Note first that if  the presence ot $T_0$ impose that $\s_i$ or $\s_j$ are fixed, that is when $i$ or $j$ are in
${\rm sf}(T_0)$, see \eqv(def:sspT-a),  then the corresponding truncated correlation
is zero. On the other hand, assuming that $|i-j|\ge 2$,
depending if  $|i-j|$ is smaller or larger than $C$, the dominant terms in \eqv(pets1) coming from the
cluster expansion of $\log  {\widetilde Z^{++}_\L(T_0,r+t(i,j))}$ are different.
In the first case it comes from a single polymer made of a single contour with two unit triangles located at site $i$ and $j$, say $T_i,T_j$
that should be compatible with the presence of $T_0$.
In the second case it comes from a single polymer made of two contours each one made of an unit triangle located at site
$i$ and $j$, that should be compatible with the presence of $T_0$ .

It follows from [\rcite{CMPR}], appendix 2, and proposition 5.4 there,  that the corresponding activity is
$$
\xi^{\overline \s(T_0)}(i)\xi^{\overline \s(T_0)}(j)e^{-\b (r+t_i) 2\overline \s_i(T_0)} e^{-\b (r+t_j) 2\overline \s_j(T_0)}
\left[e^{-\b \left[H^{++} _{T_0}(T_i,T_j)-H^{++}_{T_0}(T_i)-H^{++}_{T_0}(T_j)\right]} -1\right]
\Eq(acti42)
$$
where we have used \eqv(excess).
It can be checked by simple algebra that
$$
H^{++} _{T_0}(T_i,T_j)-H^{++}_{T_0}(T_i)-H^{++}_{T_0}(T_j)=-\,
\frac{2{\overline \s}_i(T_0){\overline  \s}_j(T_0)
}{|i-j|^{2-\a}}.
\EQ(acti43)
$$
 Therefore, taking into account  the error terms that come from the cluster expansion, we get
$$
\eqalign{
&\mu^{+}_\L(r)[\s_i,\s_j| \SS_{T_0}]=\cr
&\xi^{\overline \s(T_0)}(i)\xi^{\overline \s(T_0)}(j)e^{-\b r 2\overline \s_i(T_0)} e^{-\b r 2\overline \s_j(T_0)}
4\overline \s_i(T_0)\overline \s_j(T_0)\left[e^{\frac{2{\overline \s}_i(T_0){\overline  \s}_j(T_0)
}{|i-j|^{2-\a}}}-1\right]
\left(1 \pm e^{-\frac{\b}{64}(\frac{\z_\a}{\a(1-\a)}-3\d)}\right)\cr
}\EQ(tr42)
$$
inserting it in \eqv(5bcebis), it  implies \eqv(lpc1). \eop

\medskip
\noindent{ \bf Acknowledgements:} {\it We are indebted to Errico Presutti for many clarifying   discussions, interest and
suggestions. }

\medskip

  \centerline{\bf References}
\vskip.3truecm

\item{[\rtag{ACCN}]} Aizenman, M., Chayes, J.,  Chayes, L. and Newman, C.,:
{Discontinuity of the magnetization in one--dimensio\-nal $1/|x-y|^2$ percolation, Ising and Potts models.}
{\it J. Stat. Phys.} {\bf 50} no. 1-2 1--40 (1988).

 \item{[\rtag{BLP1}]} J. Bricmont, J. Lebowitz and  C. E. Pfister:
 On the equivalence of boundary conditions
  {\it J. Stat. Phys.} {\bf 21} 573--582 (1979)

 \item{[\rtag{BFP}]} R. Bissacot, R. Fern\'andez  and A. Procacci:
 On the Convergence of Cluster Expansions for Polymer
Gases
 {\it J.  Stat. Phys} {\bf 139} 598--617 (2010)

  \item{[\rtag{BS}]} S.E. Burkov and Ya. G. Sinai:
  Phase diagrams of one--dimensional lattice models with long--range antiferromagnetic interaction
  {\it Russian Math Survey} {\bf 38} Vol 4, 235--257 (1983)

 \item{[\rtag {CFMP}]} M. Cassandro, P.  A. Ferrari, I. Merola and E. Presutti:
{ Geometry of contours and Peierls estimates in $d=1$ Ising models with long range interaction.}
{\it  J. Math. Phys.} {\bf 46}, no 5,    (2005)
\item{[\rtag{CO}]}
M. Cassandro and  E. Olivieri:   Renormalization group and analyticity in one dimension:
a proof of Dobrushin's theorem. {\it Commun. Math. Phys} {\bf 80} 255--270 (1981).

 \item{[\rtag{CMPR}]} M. Cassandro, I. Merola , P.Picco, and U. Rozikov:
{One-Dimensional Ising Models with Long Range
Interactions: Cluster Expansion, Phase-Separating Point}
 {\it Comm. Math.Phys.} {\bf 327} No 3, 951-99, 1 (2015).

 \item{[\rtag{COP1}]} M. Cassandro, E. Orlandi, and P.Picco:
{ Phase Transition in the 1d Random Field   Ising Model  with  long range interaction. }
{\it   Comm. Math. Phy.}, {\bf 288},  731-744 (2009)

\item{[\rtag{COP2}]} M. Cassandro, E. Orlandi, and P.Picco:
{Typical Gibbs configurations for the  1d Random Field   Ising Model  with  long range interaction. }
{\it   Comm. Math. Phy.}, {\bf 309} 229-253 (2012)

\item{[\rtag{D0}]} R. Dobrushin:
The description of a random field by means of conditional probabilities and. conditions of its regularity.
{\it  Theory Probability Appl.} {\bf 13}, 197-224 (1968)
\item{[\rtag{D1}]} R. Dobrushin,
The conditions of absence of phase transitions in one-dimensional classical systems:
{\it Matem. Sbornik}, {\bf 93} , N1, 29-49 (1974)

\item{[\rtag{D2}]} R. Dobrushin:
Analyticity of correlation functions in one-dimensional classical systems with slowly decreasing potentials.
{\it Comm. Math. Phys.} {\bf 32} , N4, 269-289, (1973)

\item{[\rtag{D3}]} R. Dobrushin:
Gibbs State Describing Coexistence of Phases for a Three-Dimensional Ising Model
 {\it Theory Probab. Appl.} {\bf 17} ,582-600 (1972)

 \item{[\rtag{DKS}]} R. Dobrushin, R. Koteck\'y, S. Shlosman:
 {\it Wulff construction. A global shape from local interaction.}
Translations of Mathematical Monographs, 104. American Mathematical Society, Providence, RI, 1992

\item{[\rtag{DS}]}  Dobrushin, R.L. and Shlosman, S.: Large and moderate deviations in the Ising model. {\it Adv. in Soviet
Math.} {\bf 20} , 91-220 (1994).

  \item{[\rtag{Dy1}]} F.J. Dyson:
 {  Existence of  phase transition in a  one-dimensional Ising ferromagnetic.}
{ \it Comm. Math. Phys.},{\bf 12},91--107,  (1969).

    \item{[\rtag{Dy2}]} F.J. Dyson:
    Non-Existence of Spontaneous Magnetization
in a One-Dimensional Ising Ferromagnet \break
    { \it Comm. Math. Phys.},{\bf 12}, 212--215 (1969)

      \item{[\rtag{Dy3}]} F.J. Dyson:
An Ising Ferromagnet with Discontinuous Long-Range Order
{ \it Comm. Math. Phys.},{\bf 21} 269--283 (1971)

\item{[\rtag{E}]} R. Ellis: {\it Entropy, large deviations and Statistical Mechanics}
New York, Springer (1988)

  \item{[\rtag{FVV}]}M.  Fannes, P. Vanheuverzwijn and A.  Verbeure:
Energy--Entropy Inequalities for Classical Lattice Systems
 {\it J. Stat. Phys.} {\bf 29} No 3 547--560 (1982)

\item{[\rtag{FKG}]} C. Fortuin, P. Kasteleyn, J. Ginibre:
{Correlation inequalities on some partially ordered set}
{\it Comm. Math. Phys.} {\bf 22} 89-103 (1971).

\item{[\rtag{FS}]} J.  Fr\"ohlich and  T. Spencer:
 {  The phase transition in the one-dimensional Ising model with
 $\frac 1 {r^2}$ interaction energy.} { \it Comm. Math. Phys.}, {\bf 84}, 87--101,   (1982).

\item{[\rtag{GaMi}]}  G. Gallavotti and S. Miracle Sol\'e:
Statistical mechanics of lattice systems
{\it Comm. Math. Phys.} {\bf 5}, 317-323 (1967)

\item{[\rtag {GMM}]} G. Gallavotti, A. Martin-L\"of, S. Miracle-Sol\'e:
Some problems connected with
the description of coexisting phases at low temperatures in
Ising models, in
"Mathematical Methods in Statistical Mechanics", A. Lenard, ed., pp. 162-202, Springer, Berlin, (1973).

\item{[\rtag{Hi}]}  Y. Higuchi:
On some Limit Theorems Related to the Phase Separation Line in the Two-dimensional
Ising Model.
{\it  Z. Wahrscheinlichkeitstheorie verw. Gebiete.}  {\bf 50}, 287Ð315 (1979)

\item{[\rtag{I}]} J.Z. Imbrie:
{ Decay of correlations in the one-dimensional Ising model with $J_{ij}=\mid i-j\mid^{-2}$,}
{\it Comm. Math. Phys.} {\bf  85}, 491--515. (1982).

\item{[\rtag{IN}]} J.Z. Imbrie and C.M. Newman:
{ An intermediate phase with slow decay of correlations in one-dimensional
 $1/\vert x-y\vert ^2$ percolation, Ising and Potts models.}
 {\it  Comm. Math. Phys.}  {\bf 118} , 303--336 (1988)

\item{[\rtag{Io1}]} Ioffe, D.: Large deviations for the 2D Ising model: A lower bound without cluster expansions.
{\it J. Stat. Phys} . {\bf 74}, 411-432 (1994)

\item{[\rtag{Io2}]} Ioffe, D.: Exact large deviations bounds up to
$T_c$  for the Ising model in two dimensions. {\it Prob. Th.
Rel. Fields}  {\bf 102} , 313-330 (1995)

 \item{[\rtag{LL}]} E.H. Lieb and M. Loss:
 {\it Analysis} Graduate text in Mathematics, vol 14, American Mathematical Society, Providence Rhode Island (2001).

 \item{[\rtag{MS}]}  R.A. Minlos, Ya.G. Sinai:
  The phenomenon of phase separation at low
temperatures in certain lattice models of a gas, I and II, Math. USSR Sbornik
2, 339--395 (1967) and Trans. Moscow Math. Soc. 19, 121--196 (1968)

\item{[\rtag{OV}]} E. Olivieri, M.E. Vares:
{\it Large Deviations and Metastability}
Encyclopedia of Mathematics and its applications, Vol 100, Cambridge University Press

\item {[\rtag{PF}]}  Ch.-E. Pfister:
 Large deviations and phase separation in the two-dimensional Ising
model,
  {\it Helv. Phys. Acta} {\bf  64} , no. 7, 953-1054 (1991).
\item{[\rtag{PFVe1}]}  C.-E. Pfister, Y Velenik,: Large deviations and continuum limit in the 2D Ising model ,
{\it   Probab. Theory  Related  Fields}  109, 435-506 (1997).

 \item{[\rtag {PS}]} A. Procacci and B. Scoppola:
 Polymer gas Approach to $N$--body lattice systems
 {\it J. Stat. Phys.} {\bf 96} 49--68 (1999)

\item {[\rtag{RT}]} J. B.  Rogers and C.J. Thompson:
Absence of long range order in one dimensional spin systems.
{\it  J. Stat. Phys.} {\bf 25}, 669--678 (1981)

\item {[\rtag{Ru}]} D. Ruelle:
Statistical mechanics of one-dimensional Lattice gas,
{\it  Comm. Math. Phys.} {\bf 9}, 267--278 (1968)

\item{[\rtag{Th}]} D. J. Thouless:
    Long-Range Order in One-Dimensional Ising Systems,
{\it  Phys. Rev.} {\bf 187}, 732-733 (1969)

\item{[\rtag{W}]}  Wulff, G.: Zur Frage der Geschwindigkeit des Wachstums und der Auf\"osung der Kristallfl\"achen.
{\it Zeit\-schrift f\"ur Kristallographie}
{\bf 34}, 449-530 (1901)

\vfill
\eject

\end